\documentclass{aa}
\usepackage{natbib}
\usepackage{hyperref}

\usepackage{graphicx, graphics}
\usepackage{CJK}
\usepackage{bbm}
\usepackage{amsmath, amssymb, bm}
\usepackage[nodayofweek]{datetime}
\usepackage{txfonts}
\usepackage{ulem}
\usepackage{booktabs}  
\usepackage[switch]{lineno}
\usepackage{threeparttable}
\usepackage{longtable}
\usepackage{adjustbox}

\usepackage{tikz}
\definecolor{lime}{HTML}{A6CE39}

\newdateformat{monthyearday}{%
  \THEYEAR\ \monthname[\THEMONTH] \twodigit{\THEDAY}}
\newdateformat{daymonthyear}{%
  \twodigit{\THEDAY}\ \monthname[\THEMONTH] \THEYEAR}

\begin{document} 

   \title{Fates of the sub-stellar objects (FOSSO)}
   \subtitle{I. Fates of the known brown dwarfs in main-sequence--BD binaries}
   
   \author{Zhangliang Chen\inst{\ref{inst-sysu}, \ref{inst-csst}}
   \and
   Hongming Jin\inst{\ref{inst-sysu}, \ref{inst-csst}}
   \and
    Hongwei Ge\inst{\ref{inst-YNO}, \ref{inst-ICS}, \ref{inst-CAS}}
   \and 
    Cong Yu\inst{\ref{inst-sysu}, \ref{inst-csst}}
    \and 
    Kejun Wang\inst{\ref{inst-sysu}, \ref{inst-csst}}
    \and  
   Dichang Chen\inst{\ref{inst-sysu}, \ref{inst-csst}}\fnmsep\thanks{Corresponding author}
    \and
   Bo Ma\inst{\ref{inst-sysu}, \ref{inst-csst}}  \fnmsep\thanks{Corresponding author} 
          }

   \institute{School of Physics and Astronomy, Sun Yat-sen University,
Zhuhai 519082, People's Republic of China;    \email{\url{chendch28@mail.sysu.edu.cn}, \url{mabo8@mail.sysu.edu.cn}}  \label{inst-sysu} %
         \and
         CSST Science Center for the Guangdong-HongKong-Macau Great Bay Area, Sun Yat-sen University, Zhuhai 519082, People's Republic of China \label{inst-csst}
         \and
         Yunnan Observatories, Chinese Academy of Sciences,
         Kunming, 650216, People's Republic of China        \label{inst-YNO}
         \and
         International Centre of Supernovae, Yunnan Key Laboratory, 
         Kunming 650216, People's Republic of China         \label{inst-ICS}
         \and
         University of Chinese Academy of Sciences, 
         Beijing 100049, People's Republic of China         \label{inst-CAS}
             }

   \date{Received ; accepted }

\abstract
   {
  Understanding the survival and orbital evolution of brown dwarf (BD) companions during the post-main-sequence (MS) evolution of their host stars is increasingly important, especially with recent discoveries of many substellar companions around white dwarfs (WDs).
  }
  {
  We investigate the long-term evolution and final outcomes of BDs orbiting low-mass MS stars as these evolve into WDs. By comparing forward-modeling populations with observed WD--BD binaries, we test evolutionary models and predict the existence of yet-undetected systems.
  }
  {
  We employ the COMPAS binary population synthesis code to evolve observed MS--BD systems through the post-MS phases of their host stars into the WD stage, tracking orbital changes driven by mass loss, tides, and common-envelope (CE) evolution.
  }
  {
   Our simulations reproduce a period gap in the distribution of detached WD--BD binaries, consistent with observations. We also identify a boundary separating detached and semi-detached systems on the period-mass diagram, located at orbital periods of $\sim$1--2 hours depending on the BD mass.
  }
  {
    We predict that a subset of currently known MS--BD binaries will survive post-MS evolution and emerge as detached WD--BD systems, while others will undergo CE evolution and potentially form cataclysmic variables with BD donors. Our results reproduce the observed period gap in WD--BD binaries and provide quantitative predictions for the role of CE efficiency in shaping their distribution. 
    This work predicts that many WD--BD systems remain undetected, motivating targeted searches with microlensing and high-contrast imaging techniques using next-generation large telescopes.
  }


\authorrunning{Chen et al.}
\titlerunning{The fate of MS--BD binaries}

   \maketitle
%

\section{Introduction}
Brown dwarfs (BDs) are substellar objects that occupy the mass regime between giant planets and low-mass stars, typically defined as having masses between 13 and 75 Jupiter masses ($M_{\rm Jup}$; \citealt{Burrows01}). Unlike stars, BDs lack the sufficient mass to sustain stable hydrogen fusion in their cores, thus never reaching the main sequence. The detection of thousands of BDs, including a significant fraction in binary systems \citep{Feng22}, has provided invaluable data for understanding the processes of both planetary and stellar formation. A notable feature in the population statistics of BDs orbiting FGK-type main-sequence (MS) stars is the "brown dwarf desert" (BDD), characterized by a paucity of intermediate-mass BDs in close orbits \citep{Halbwach00,Marcy00,Troup16,Shahaf19}. 
This deficit suggests distinct formation mechanisms for BDs of different masses \citep{Grether06,Ma14}, a hypothesis further supported by analyses of orbital and stellar property distributions. For example, using a compiled sample from radial velocity (RV), transit and astrometry observation, \citet{Ma14} proposed a transition mass around $42.5~M_{\rm Jup}$ dividing the different eccentricity distributions for lower-mass BDs and higher-mass BDs. More recently, \citet{Giacalone25} compiled a long-period BD sample from the RV observation of California Legacy Survey, and found a metallicity transition mass around $27~M_{\rm Jup}$ that divides the metal-rich hosts of lower-mass companions and metal-poor hosts of higher-mass companions. While some studies on short-period transiting BD samples do not support such transition between planetary and stellar formation \citep[e.g.,][]{Vowell25}, suggesting that additional observations are needed to confirm this picture.

The long-term dynamical evolution of substellar companions, particularly BDs and planets, around FGK-type stars is a critical area of astrophysical research. These host stars will eventually evolve off the MS, passing through giant phases before becoming white dwarfs (WDs). During this post-MS evolution, physical processes such as stellar mass loss, tidal interactions, and stellar wind accretion can profoundly alter the orbital parameters and even the masses of their companions \citep{Nordhaus10, Veras11}. While the survivability and dynamical fates of planetary companions have been extensively studied \citep{Villaver09, Veras11, Nordhaus13, Mustill14}, investigations specifically dedicated to brown dwarfs in this context remain comparatively less explored.

Brown dwarfs, owing to their significantly larger masses compared to giant planets, are more likely to survive the common-envelope (CE) phase, thereby facilitating the formation of close WD--BD binary systems \citep{Nordhaus13}. Their greater mass also leads to stronger tidal interactions with their evolving host stars. Studies of BD evolution during the post-MS stage are therefore crucial for constraining poorly understood binary interaction parameters, such as those governing the CE phase, and for testing theories related to the BDD \citep{Zorotovic22, ChenZL24}. Despite the identification of more than a thousand BD candidates orbiting MS stars \citep{Gaia23a,Kiefer25}, only a limited number of BDs have been confirmed orbiting WDs \citep[e.g.,][]{Casewell18EPIC}. This observational scarcity is primarily due to the intrinsic faintness of BDs, making their detection challenging in WD systems. Nevertheless, population synthesis models predict a substantial, yet largely unobserved, population of WD--BD binaries that should have evolved from MS--BD progenitors over gigayear timescales \citep{Zorotovic22, ChenZL24}.

This paper, the first work of project Fates of the Known Sub-stellar Objects (hereafter as FOSSO I) addresses this gap by investigating the long-term evolution of a comprehensive sample of observed brown dwarf companions to A-, F-, G-, and K-type stars. We employ detailed numerical simulations to trace their evolution through the post-main-sequence phases of their host stars into the white dwarf stage. Our primary objectives are to: (1) predict the ultimate fates and orbital distributions of these BDs, (2) compare our simulated WD--BD populations with the currently observed sample to validate and refine formational and evolutionary models, and (3) provide guidance for future observational campaigns aimed at discovering new WD--BD systems. By focusing on the distinct post-MS evolutionary behavior of brown dwarfs, this study aims to bridge the current understanding between planetary and stellar companion evolution, thereby contributing to a more complete picture of substellar populations and their role in binary stellar evolution.

The structure of this paper is as follows: Section \ref{sec:data} describes the observational datasets of MS--BD and WD--BD binary systems used in our analysis. Section \ref{sec:method} details our numerical methods and presents the simulation results. A comprehensive discussion of our findings is provided in Section \ref{sec:disc}, followed by our conclusions in Section \ref{sec:conc}.

\section{Data and observation} \label{sec:data}
We have gathered data of 196 known MS--BD binaries from the literature in an attempt to investigate the future fate and distribution of these systems.
In Section~\ref{sec:method} we will demonstrate that most of these systems will evolve either to close WD--BD binaries or to wide WD--BD binaries in the future.
In addition, we have also collected information of 22 known WD--BD binaries from the literature, in an attempt to compare them with the future evolution of known MS--BD binaries.

\subsection{The BD companion around MS star} \label{sec:data1}
\begin{figure}[ht!]
\centering
\includegraphics[scale=0.375]{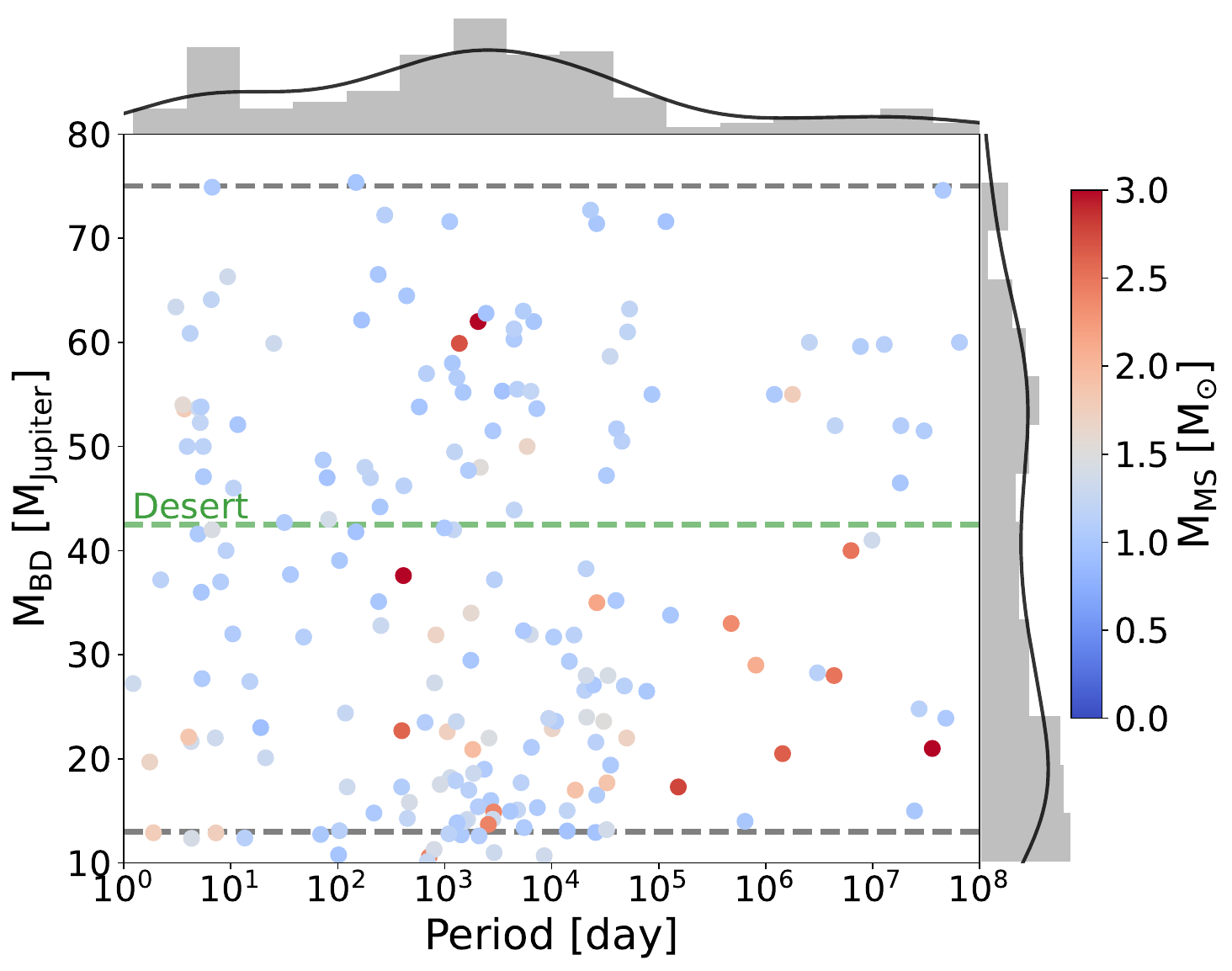}
\caption{
Mass–period distribution of the observed MS--BD binary systems used in this study. The color scale indicates the mass of the host MS star.
The top and right histograms with kernel density estimate (KDE) show the distributions of the orbital period and BD mass, respectively.
The grey dashed lines show the mass range of BDs, while the green dashed line represents the BD desert transition mass of $ 42.5~M_{\rm Jup}$ in \citet{Ma14}.
}
\label{fig:ob_MSBD_nofate}
\end{figure}

The observational data for MS--BD binaries used in this study were primarily obtained from recent literature, including \citet{Rosenthal21,Grieves21,Feng22,Stevenson23,Rothermich24,Kiefer25,Vowell25}, publicly accessible astronomical databases \citep{Exoplaent_eu, Exoplanet_org}, and references therein.
We performed cross-checking between catalogs and applied the following selection criteria:\\
1) the BD has mass measurement in the mass range between 13--75~$M_{\rm Jup}$;\\
2) the mass of the host star is between 0.9--8.0~$M_{\odot}$, to ensure it can evolve to WD in 13.7 billion years;\\    
3) the metallicity $Z$ of the host star is between 0.001--0.03;\\ 
4) the host star is in its MS evolution stage;\\
5) the orbital period is less than $10^8$~days.\\   
Based on these criteria, we selected 196 MS--BD binary systems. The complete catalog, including the system parameters and corresponding references, is provided in Appendix~\ref{apx:MSBD_catalog}.
The BD mass-orbital period distribution of this sample is shown in Figure~\ref{fig:ob_MSBD_nofate}, where the color represents the MS mass.
The top and right histograms show the distributions of the orbital period and BD mass, respectively. We also calculate and plot the kernel density estimate (KDE) of the distributions.
In this sample, the majority of the MS stars are solar-type stars, with masses around 1~$M_{\odot}$. Their expected WD remnants cover two mass ranges: systems that experience CE evolution may produce lower-mass WDs \citep[0.4--0.5~$M_{\odot}$; e.g.,][]{Zorotovic22}, while those that avoid CE evolution would follow single star evolution model, and produce typical-mass WDs \citep[0.5--0.6~$M_{\odot}$; ][]{Hurley00, Hurley02}.
The BD companions show a relatively uniform mass distribution, with a shallow gap around the transition mass of $42.5~M_{\rm Jup}$ proposed by \citet{Ma14}.
The orbital periods are mainly concentrated in the range of $10^3$ to $10^4$~days, and show a gap around $10^5$~days likely arising from observational biases in different methods.

\subsection{The BD companion around WD}
To date, only 11 detached close ($P_{\rm orb}<12$~hour) WD--BD binary systems have been identified:    
SDSS J1411+2009 \citep[WD+T5;][]{Beuermann13, Littlefair14};
SDSS J1205-0242 \citep[WD+L0;][]{Parsons17, Rappaport17};
WD 1032+011 \citep[WD+L5;][]{Casewell20};
ZTF J0038+2030 \citep[WD+BD;][]{Roestel21};
WD0137-349 \citep[WD+L8;][]{Maxted06, Burleigh06};
NLTT 5306 \citep[WD+dL5;][]{Steele13};
SDSS J1557+0916 \citep[WD+L3--L5;][]{Farihi17};
EPIC 212235321 \citep[WD+L3;][]{Casewell18EPIC};
GD 1400 \citep[WD+L6--L7;][]{Farihi04, Dobbie05, Burleigh11, Casewell24};
WD 0837+185 \citep[WD+T8;][]{Casewell12};
SDSS J1231+0041 \citep[WD+M/L;][]{Parsons17}.
And 11 wide ($P_{\rm orb}>100~\rm yr$) WD--BD binary systems have been identified:
KMT-2020-BLG-0414 \citep[][]{Zhang24NatAs};
PHL 5038 \citep[WD+L8;][]{Steele09};
GD 165 \citep[WD+L4;][]{Becklin88};
SDSS J2225+0016 \citep[WD+L4;][]{French23};
LSPM J0055+5948 \citep[WD+T8;][]{Meisner20};
COCONUTS-1 \citep[WD+T4;][]{Zhang20};
VVV J1256-6202 \citep[WD+L3;][]{ZhangZH24VVV};
Wolf 1130 C \citep[][]{Mace13, Mace18, Burgasser25};
WD 0806-661 \citep[WD+Y1;][]{Luhman11};
LSPM J1459+0851 \citep[WD+T4.5;][]{Day-Jones11};
LSPM J0241+2553 \citep[][]{Deacon14}.
Among these, nine close systems and ten wide systems possess reliable and precise measurement data.
The system parameters of these 19 systems are summarized in Table~\ref{tab:ob_WDBD}.

\begin{table}[ht!]
        \centering
        \caption{ Parameters of WD$-$BD binaries from observations. }
        \label{tab:ob_WDBD}
        \resizebox{0.48\textwidth}{!}{%
        \begin{tabular}{lccc} 
                \hline
                     & M$_{\rm WD}$ & M$_{\rm BD}$ & Period \\
        Close system & [$M_{\odot}]$ & [$M_{\odot}$] & [min] \\
                \hline
        EPIC~212235321 & $0.47\pm0.01$ & $0.055^{+0.007}_{-0.010}$ & 68.2\\
        SDSS~J1205$-$0242 & 0.39$\pm0.02$ & $0.049\pm0.006$ & 71.2\\
        NLTT~5306 & $0.44\pm0.04$ & $0.053\pm0.003$ & 101.9\\
        WD~0137$-$349 & $0.39\pm0.035$ & $0.053\pm0.006$ & 115.6\\
        SDSS~J1411$+$2009 & $0.53\pm0.03$ & $0.050\pm0.002$ & 121.7\\
		WD~1032$+$011 & $0.45\pm0.05$ & $0.067\pm0.006$ & 131.8\\
        SDSS~J1557$+$0916 & $0.447\pm0.043$ & $0.063\pm0.002$ & 136.4\\
        GD~1400 & $0.68\pm0.03$ & $0.074\pm0.006$ & 598.8\\
		ZTF~J0038$+$2030 & $0.50\pm0.02$ & $0.0593\pm0.004$ & 622.0\\
        \hline
        Wide system & [$M_{\odot}]$ & [$M_{\odot}$] & [yr] \\
        \hline
        KMT-2020-BLG-0414 & $0.49^{+0.06}_{-0.03}$ & $0.0269^{+0.004}_{-0.003}$ & 143.9\\
        PHL 5038 & $0.53\pm0.02$ & $0.069\pm0.001$ & 700.0\\
        GD 165 & $0.64\pm0.02$ & $0.060\pm0.015$ & 1628.4\\
        SDSS J2225+0016 & $0.66^{+0.11}_{-0.06}$ & 0.037$\pm$0.013 & 3577.8\\
        LSPM J0055+5948 & $0.474\pm0.01$ & $0.053\pm0.009$ & 11105.7\\
        COCONUTS-1 & 0.548 & $0.066^{+0.002}_{-0.003}$ & 58406.6\\
        VVV J1256-6202 & $0.62\pm0.04$ & $0.082\pm0.001$ & 60535.7\\
        Wolf 1130 & $1.24^{+0.19}_{-0.15}$ & $0.050$ & 155652.0\\
        WD 0806-661  & $0.62\pm0.03$ & 0.007 & 158384.0\\
        LSPM J1459+0851 & 0.585 & $0.066\pm0.006$ & 3892793.1\\
                \hline
        \end{tabular} 
        }
\end{table}

Recent Hubble Space Telescope (HST) studies of BD atmospheres continue to refine our understanding of cloud structure, spectral behavior, and atmospheric dynamics across the L-T-Y sequence. Time-resolved near-infrared spectroscopy with HST/Wide Field Camera 3 (WFC3) has shown that the variability in L/T-transition objects is primarily driven by heterogeneous cloud thickness and large-scale atmospheric circulation rather than simple clearing \citep{Apai13, Buenzli14}. The HST imaging programs targeting T8--Y1 dwarfs have produced important constraints on binarity and system architecture \citep{Fontanive23}. For WD--BD systems like GD~1400, HST observations have also provided complementary constraints on their BD atmosphere, revealing cloud coverage, day-night temperature contrasts, and efficient heat re-distribution \citep{Amaro25}. These studies highlight the role of HST in probing the physical and dynamical characteristics of BDs.

As shown in Table~\ref{tab:def}, throughout this paper, we focus on four sets of samples. Except for the 19 WD--BD binaries from observations, the other two samples are simulated WD--BD binaries, which are derived from the observed MS--BD binaries using the binary evolutionary models discussed in Section~\ref{sec:method1}. Here and throughout the paper, the term `simulation' or `simulated system' refers to these samples. 

\begin{table}[ht!]
        \centering
        \caption{The WD--BD binary samples. }
        \label{tab:def}
        \resizebox{0.45\textwidth}{!}{
        \begin{threeparttable}
            \begin{tabular}{lccc}
            \hline
            Sample & Type & Size \\
            \hline
            Observation &  & \\
            \hline
            Sample~1 & Close binary (post-CE, $P_{\rm orb}<12~h$) & 9 \\
            Sample~2 & Wide binary (no CE, $P_{\rm orb}>100~\rm yr$) & 10 \\
            \hline
            Simulation$^*$  &  & \\
            \hline
            Sample~3 & Close binary (post-CE, $P_{\rm orb}<12~h$) & 11 \\
            Sample~4 & Wide binary (no CE, $P_{\rm orb}>100~\rm yr$) & 94 \\
            \hline
            \end{tabular} 
        \begin{tablenotes}
            \footnotesize
            \item[*] The term `simulation' used hereafter refers to the evolutionary simulation of the observed MS--BD sample using COMPAS.
        \end{tablenotes}
        \end{threeparttable}
        }
\end{table}


\section{Method and result} \label{sec:method}
\subsection{Evolution of the MS--BD binary} \label{sec:method1}
To investigate the fate of the observed MS--BD binary systems in the post-MS stage of the primary star, we have employed the COMPAS code \citep{Riley22teamCOMPAS} to simulate their evolution over the timescale of the age of the universe. 
Based on the SSE code \citep{Hurley00} and BSE code \citep{Hurley02}, COMPAS is designed for fast population synthesis of binary evolution. We have modified the code to calculate the evolution of MS--BD binary \citep{ChenZL24}.

In the post-MS stage of the primary, the orbital evolution of a binary system is a variable-mass two-body problem \citep{veras16}.
During this stage, two factors significantly affect the companion's orbital evolution, including mass loss (ML) and tidal interactions between the two stars in the binary \citep{Nordhaus13}.
The mass loss of the primary is mostly attributed to the rapid stellar wind loss process.
In this case, the timescale of primary mass loss is much larger than the orbital period, a scenario known as the $Jeans$ $mode$ \citep{Huang63}. 
Considering the conservation of angular momentum in the system, the evolution of the orbital semi-major axis $a$ can be calculated by:
\begin{equation}\label{eq:ML}
(\frac{da}{dt})_{\rm ML} = -\frac{a}{M_{\rm tot}} \frac{dM_{\rm tot}}{dt},
\end{equation}
where $M_{\rm tot} = M_{\rm MS} + M_{\rm BD}$ is the total mass of the binary.
When the primary evolves off the MS and reaches the giant branch stage, it undergoes significant radius expansion and mass loss. 
As a result, its companion will move to a wider orbit, while the eccentricity remains unchanged.
Moreover, stellar tidal interactions can transfer orbital angular momentum to the star's spin angular momentum, leading to orbital decay, circularization, and spin-up of the primary star.
In our simulation, we have considered orbital evolution under both equilibrium and dynamical tides,
derived by Kapil et al. (in prep.), based on \citet{Zahn77}, which is included in the latest version of COMPAS \citep{COMPAS25}.

For the close MS--BD binaries ($P\lesssim1000~ \rm d)$, the orbital expansion caused by the mass-loss process 
may not be sufficient to prevent engulfment by the expanding primary star.
If the primary's envelope fills its Roche Lobe (RL), mass transfer can happen by RL overflow (RLOF).
The equivalent RL radius can be estimated following the formula in \citet{Eggleton83}:
\begin{equation}\label{eq:RL}
    R_{\rm MS,RL}= \frac{0.49aq^{2/3}}{0.6q^{2/3}+ \rm ln(1+\it q^{\rm 1/3})},
\end{equation}
where the $q=M_{\rm MS}/M_{\rm BD}$ is the mass ratio of the primary star to the companion.
According to \citet{Ge20b, Ge23, Ge24a}, the mass transfer of a binary with an initial mass ratio larger than a critical mass ratio tends to be unstable. 
Once the unstable mass transfer happens, the primary stellar core and the companion BD will soon be surrounded by the primary's envelope, which is known as the CE phase.
During this phase, the stellar core and the BD will keep spiraling in because of the dynamical interaction with the envelope material \citep{Livio88}, which eventually leads to the orbital decay and circularization.  
During this phase, the binary will experience orbital decay and circularization due to interactions with the envelope \citep{Livio88}. 
If the binary has enough energy to disperse the CE, it will become an extremely close post-CE binary. 
As is described in \citet{XuLi10}, the orbital decay can be calculated by:
\begin{equation}
    \frac{a_{\rm f}}{a_{\rm i}}=\frac{M_{\rm 1,c}M_{\rm BD}}{M_1}\frac{1}{M_{\rm BD}+2M_{\rm 1,e}a_{\rm i}/(\alpha_{\rm CE}\lambda R_{\rm RL})}.
\end{equation}
The $M_{\rm 1,c}$ and $M_{\rm 1,e}$ are the core mass and envelope mass of the primary star, respectively.
The $\lambda$ is the envelope binding energy parameter, for which we use the prescription from \citet{XuLi10}.
The $\alpha_{\rm CE}$ is the energy transform rate, which is estimated to be $\sim0.3$ for MS--BD binaries \citep{Zorotovic22,ChenZL24}.
After dispersing the CE, the MS--BD binaries will ultimately evolve to WD--BD binaries with close and circular orbits.

Using a population synthesis method, \citet{ChenZL24} have identified and characterized two evolutionary channels from close MS--BD binaries to post-CE WD--BD binaries, resulting in two corresponding classes of WD--BD binaries, as shown in their Figure~1 and Section~3. 
The main difference between the two channels is the evolutionary stage at which the system enters the CE phase: 
\begin{itemize}
    \item Channel~A: CE occurs during the primary's red giant branch stage, and the system is likely to form helium-core WD--BD binary.
    \item Channel~B: CE occurs during the primary's asymptotic giant branch stage, and the system is likely to form carbon-oxygen-core WD--BD binary.
\end{itemize}
For example, recent research \citep[e.g.,][]{Zhang24NatAs} suggests that the relatively wide post-CE WD--BD might form following the channel~B in \citet{ChenZL24}.
For these close WD--BD binaries, their orbits will continue to shrink due to the angular momentum loss (AML) from the gravitational wave (GW) radiation.
We use the formula from \citet{Schreiber03} to calculate the orbital decay due to GW radiation from the moment after dispersing the CE to the age of the universe.
The orbital decay driven by GW radiation can be calculated by:
\begin{equation}        \label{eq:P_of_t_GW}
    P^{8/3} = P_0^{8/3}-\frac{256}{5}(2\pi)^{8/3}\frac{G^{5/3}}{c^5}\frac{M_{\rm WD}M_{\rm BD}}{(M_{\rm WD}+M_{\rm BD})^{1/3}}t_{\rm cooling},
\end{equation}
where the $t_{\rm cooling}$ is the WD cooling time.

As suggested by \citet{Schreiber03}, a post-CE binary system containing a WD and an MS/M-dwarf companion can evolve into a cataclysmic variable (CV) due to GW-induced AML process and the expansion of the companion, and eventually start mass transfer through RLOF. 
In this traditional evolutionary channel, the donor is initially stellar and evolves towards shorter orbital periods until reaching the period minimum, after which further mass loss could lead to a partially degenerate donor and the period bounce phase \citep[e.g.,][]{Rappaport82,Knigge11,McAllister17}.
Consequently, many CVs hosting BD-mass donors are considered to have reached the substellar boundary through stable mass loss, rather than originating from WD--BD binaries \citep[e.g.,][]{McAllister17}.
In contrast, post-CE WD--BD binaries can form CVs in a different evolutionary pathway. 
Unlike MS/M-dwarf, the structure of BDs can be well described by the polytropic model with index $n=1.5$.
Their cores, supported by electron degeneracy pressure, can be regarded as uniform-density convective cores \citep{Burrows01}. 
When a BD approaches its host WD, tidal forces can cause its radius to expand and fill its RL. In this case, material from the BD will transfer to the WD via the RLOF process, potentially forming an accretion disk around the WD and eventually transforming it into a CV \citep[e.g.][]{Casewell18,Casewell18EPIC,Rappaport17}. 
Because the degenerate BD is originally formed and partially degenerate, the post-CE WD--BD binary can initiate RLOF at shorter orbital periods than typical WD--MS/M-dwarf binary.
For example, \citet{Rappaport17} have studied how the very short-period transiting BD in pre-CV binary WD~1202-024 (SDSS~J1205$-$0242) would evolve into a CV system in the near future.
The close post-CE WD--BD systems have a minimum allowed orbital period, below which the BD would fill its RL and start mass transfer. 
In our simulation, we calculate the minimum period of the post-CE WD--BD binaries following Equation~16 in \citet{Rappaport21}:
\begin{equation} \label{PminR21}
    \rm ln \mathit{P}_{\rm min, ~BD} \simeq  1.01
    - 1.085ln(\mathit{\frac{m_{\rm BD}}{M_{\odot}}})
    - 0.052ln^2(\mathit{\frac{m_{\rm BD}}{M_{\odot}}}),
\end{equation}
where the $P_{\rm min, ~BD}$ is in minutes.   
In our simulation, we found that some systems, despite having enough energy to disperse the CE, ended up with the orbital periods below the $P_{\rm min, ~BD}$. This suggests that the BDs in these systems should have been tidally disrupted during the CE phase, and eventually merge to single WDs.

While for the wide MS--BD binaries ($P\gtrsim1000$~d), except for the extremely eccentric systems ($e\gtrsim0.9$), the tidal dissipation of the primary stars has little effect on their orbits due to the large separation.
Therefore, the BD companion in these systems will not experience the CE phase. Instead, the mass loss of the primary star will drive the BD to move to a wider orbit as the primary star evolves to a WD.
Similarly, Jovian planets on wide orbits can also avoid engulfment and survive the post-MS evolution of their host stars. These objects around nearby WDs may be detectable via infrared imaging \citep{Burleigh02}. 

\subsection{Evolution Result}\label{sec:result}

\begin{figure}[ht!]
\centering
\includegraphics[scale=0.375]{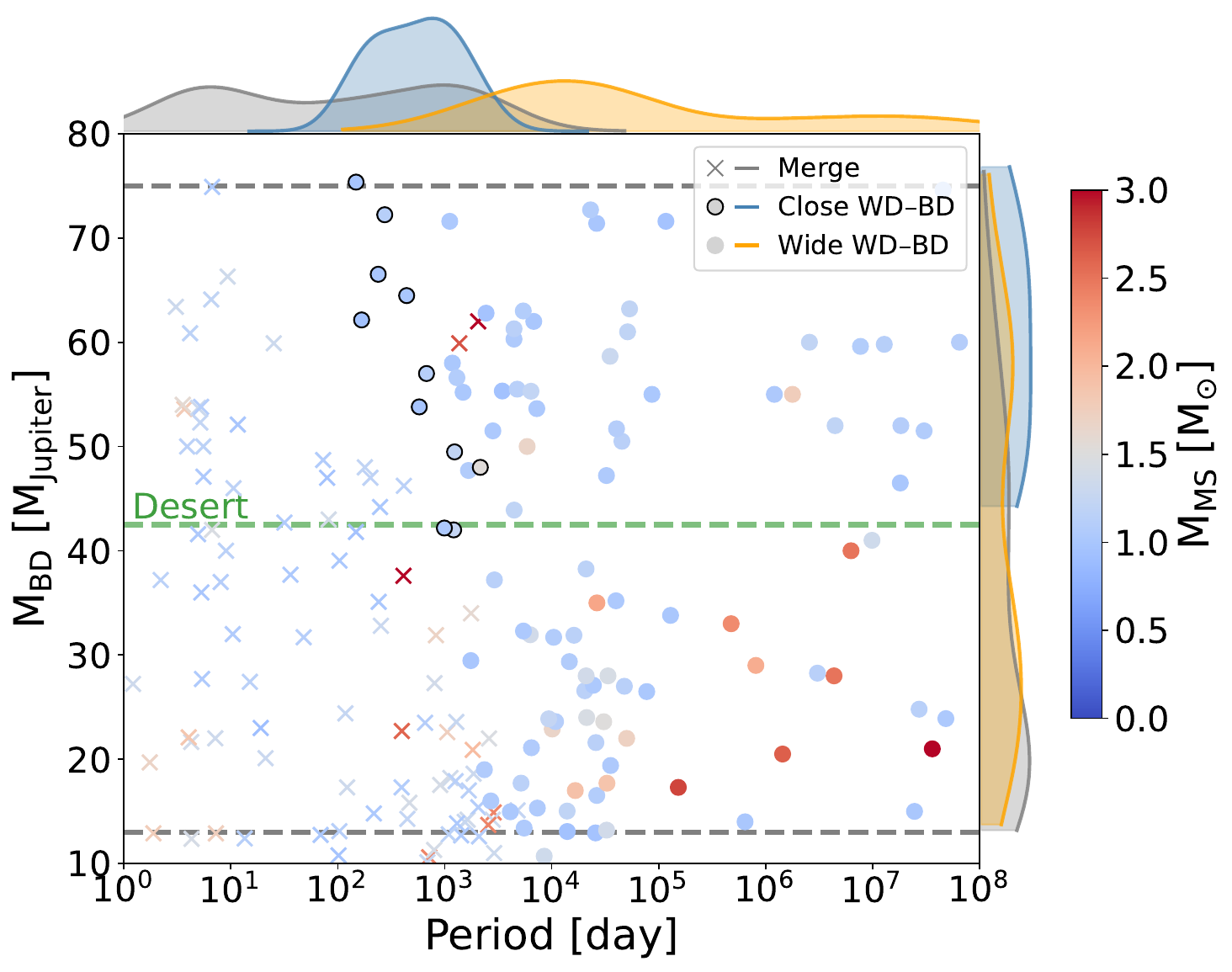}
\caption{
Same as Figure~\ref{fig:ob_MSBD_nofate}, all the markers represent the initial MS--BD systems, while different markers indicating the different fates.
The top and right KDE plots represent the distributions of different fates: grey, blue and orange correspond to merged system, close WD--BD and wide WD--BD system, respectively.
}
\label{fig:ob_MSBD}
\end{figure}

\begin{figure}[ht!]
\centering
\includegraphics[scale=0.45]{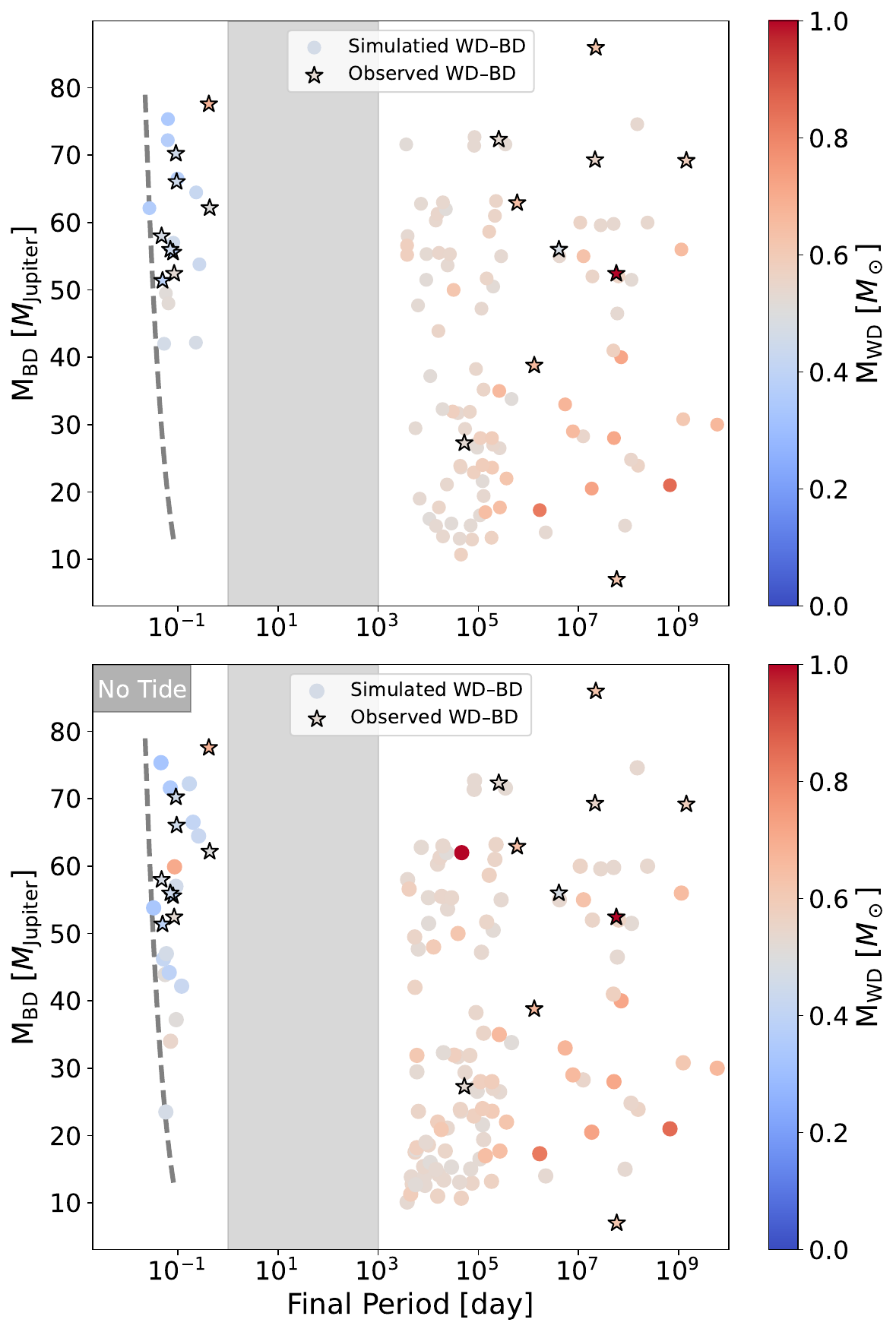}
\caption{
Period-mass distribution for the future WD--BD system of the observed MS--BD binaries, calculated with (top panel) and without (bottom panel) tidal effect. The stars mark the observed closed and wide WD--BD binaries listed in Table~\ref{tab:ob_WDBD}. The gray area shows the period gap around 1--1000~day. The dashed line represents the minimum period for the binary to evolve to CVs.
}
\label{fig:fate_distribution}
\end{figure}

Adopting the evolutionary framework described in Section~\ref{sec:method1}, we calculated the evolutionary fate of 196 MS--BD binaries selected in Section~\ref{sec:data1}. The results of our simulation are presented in Figures~\ref{fig:ob_MSBD} and \ref{fig:fate_distribution}. In our simulation, 94 binaries do not experience a CE phase and remain as wide WD--BD binaries, while 102 binaries undergo a CE phase during the GB stage of the primary star. Of the 102 systems experiencing the CE phase, 91 merge in CE and evolve into single WDs, while other 11 systems survive as detached WD--BD binaries after CE dispersal: BD+63 974~b, HD~156312~Bb, HD~115517~b, HD~30246~b, HD~87646~Ac, HD~104289~b, HD~140913~b, HD~132032~b, HD~5433~b, HD~136118~b, and KOI-415~b. The orbital parameters of these systems after CE dispersal are listed in Table~\ref{tab:fates_close_WD_BD}, including the final stellar mass, orbital period, and the timescales for the primary to evolve into a WD and to reach the CV phase, which are calculated using Equations~\ref{eq:P_of_t_GW} and \ref{PminR21}.

Figure~\ref{fig:ob_MSBD} shows the distribution of the initial parameters of the sample, with different markers indicating their different fates.
From the top and right KDE curves, we find that systems with a more massive BD and an intermediate initial orbital period (100--1000~days) are more likely to form post-CE WD--BD binaries.
Figure~\ref{fig:fate_distribution} shows the period-mass distribution of future WD--BD binary systems. The upper and lower panels correspond to results with and without tidal dissipation effects, respectively. As noted by \citet{Nordhaus13}, the combined effects of mass-loss-induced orbital expansion, tidal dissipation, and CE-ejection-induced orbital decay should create a period gap in the distribution of sub-stellar companions around their host WDs. For the future WD--BD population, we also identify a period gap of around 1 to 1000 days. Confirmed WD--BD binaries listed in Table~\ref{tab:ob_WDBD} are marked with pentagrams; they are distributed on either side of the period gap, in good agreement with our simulation results. The gray dashed line indicates the minimum allowed orbital period, below which systems are likely to become CVs. To date, no detached or semi-detached CV systems containing low-mass BDs with periods below this minimum have been observed, suggesting that most such systems should have merged during the CE phase.

\begin{table*}[ht!]
        \centering
        \caption{ Parameters of WD$-$BD binaries survive from CE in our simulation, where $\alpha_{\rm CE}=0.3$. }
        \label{tab:fates_close_WD_BD}
        \begin{tabular}{lccccccc} 
                \hline
                     & M$_{\rm BD}$ & P$_{\rm ini}$ & P$_{\rm post-CE}$ & M$_{\rm MS}$ & M$_{\rm WD}$ & $t_{\rm MS\rightarrow WD}$ & $t_{\rm WD\rightarrow CV}$ \\
        Close system & [$M_{\odot}$] & [day] & [min] & [$M_{\odot}$] & [$M_{\odot}$] & [Gyr] & [Gyr] \\
                \hline
        BD+63 974 b & 0.045 & 2149 & 93 & 1.53 & 0.52 & 2.6 & 0.9\\
        HD 156312 Bb & 0.063 & 238 & 141 & 0.99 & 0.37 & 11.7 & 3.0\\
		HD 115517 b & 0.061 & 440 & 332 & 1.0 & 0.42 & 11.3 & 28.3\\
        HD 30246 b & 0.040 & 990 & 330 & 1.03 & 0.47 & 10.6 & 39.0\\
        HD 87646 Ac & 0.054 & 675 & 117 & 1.12 & 0.44 & 6.8 & 1.8\\
        HD 104289 b & 0.047 & 1235 & 121 & 1.15 & 0.51 & 7.0 & 2.0\\
        HD 140913 b & 0.071 & 148 & 91 & 0.98 & 0.34 & 12.6 & 0.9\\
        HD 132032 b & 0.068 & 275 & 91 & 1.12 & 0.36 & 8.5 & 0.9\\
        HD 5433 b & 0.051 & 577 & 383 & 0.98 & 0.43 & 12.1 & 48.2\\
        HD 136118 b & 0.040 & 1210 & 77 & 1.24 & 0.46 & 5.0 & 0.5\\
        KOI-415 b & 0.059 & 167 & 39 & 0.94 & 0.35 & 12.1 & 0.00035\\
                \hline
         \end{tabular}
\end{table*}

\begin{figure}[ht!]
\centering
\includegraphics[scale=0.5]{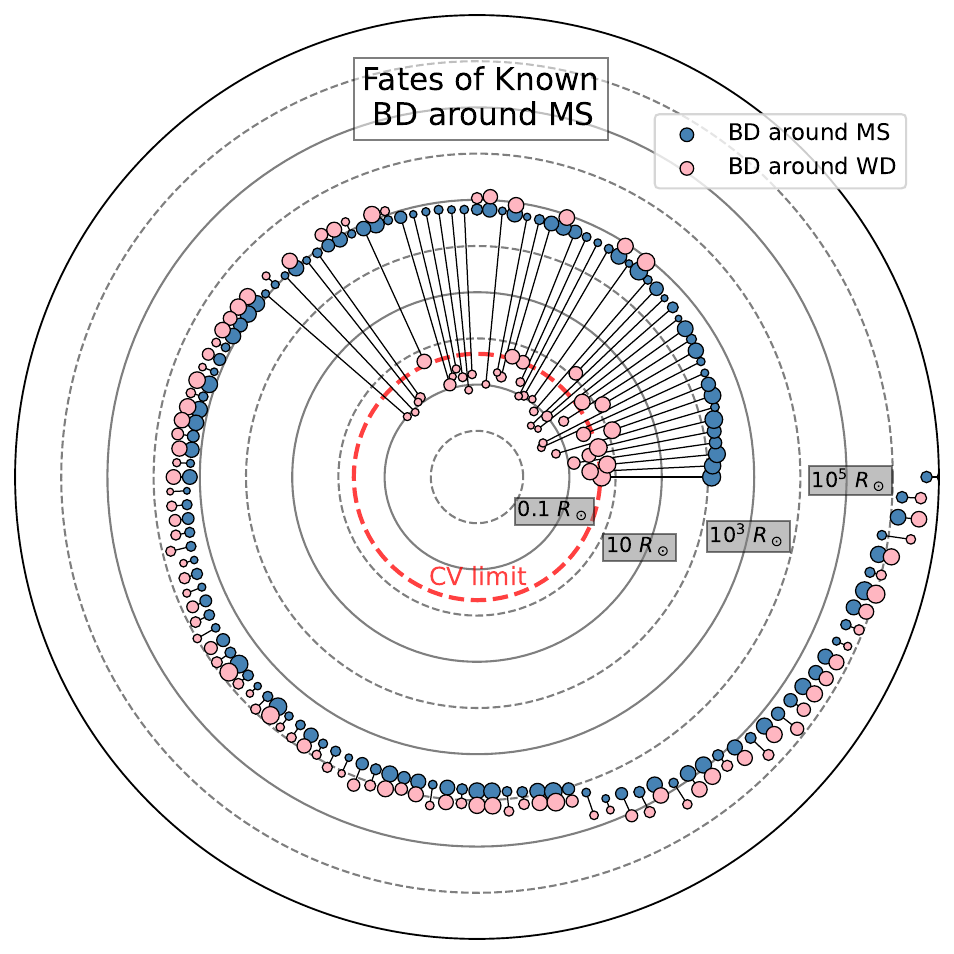}
\caption{
Fates of known BDs around MS. Blue and pink dots represent the initial and final separations of the survived MS--BD binaries. The size of the dots indicates the mass of BDs. The red dashed circle shows the average of the minimum separations for the binaries to stay detached.
}
\label{fig:fate_polar}
\end{figure}

Figure~\ref{fig:fate_polar} shows the change in the semi-major axis of MS--BD binaries that successfully evolve into WD--BD binaries. Close systems that undergo CE evolution experience significant orbital decay, whereas wider systems that avoid CE experience orbital expansion. Consequently, a separation gap appears at the WD--BD stage, roughly between 1 and 1000~$R_{\odot}$. We also plot the minimum orbital separation for a rough comparison, derived from the average Roche lobe radius of the BDs in surviving systems. Systems with separations smaller than this limit are likely to become semi-detached and enter the CV stage. Other close systems will gradually approach this limit through GW radiation emission, eventually evolving into CVs.

\subsection{Mass ratio and period distribution of WD--BD}

\begin{figure*}[ht!]
\centering
\includegraphics[scale=0.66]{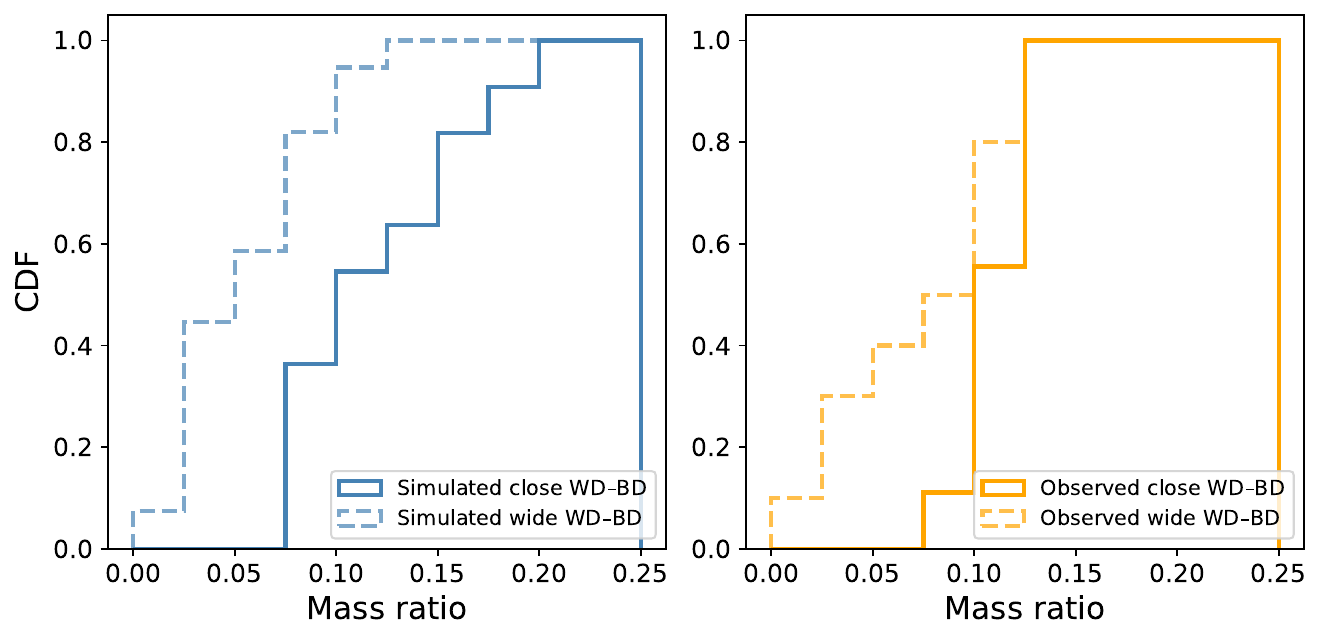}
\caption{
The CDFs of the mass ratio $q=M_{\rm BD}/M_{\rm WD}$.
The left panel shows the mass ratio distribution of close and wide simulated samples, respectively.
While the right panel shows the distribution of the observed samples.
}
\label{fig: mass_ratio}
\end{figure*}

We further perform a comparative analysis between the post-CE close and no-CE wide WD--BD binaries from simulation and observation.
As shown in Figure~\ref{fig: mass_ratio}, the two panels show the cumulative distribution functions (CDFs) of the mass ratio $q=M_{\rm BD}/M_{\rm WD}$ for the simulated and observed WD--BD binaries, with solid and dashed lines representing the close and wide systems, respectively.
From the simulation result, we find that the close systems have notably higher mass ratios than the wide systems.
This can be explained by the CE evolutionary models: CE-dispersal process can cut off the growth of primary's core, leading to a less massive WD. At the same time, the heavier BDs are more likely to survive the CE phase, resulting in higher mass ratios for close systems.
As shown in the left panel, the close binary systems in the observational sample exhibit relatively high mass ratios, consistent with the simulation results. However, the current sample size remains small, and future observations are needed to confirm this feature.

\begin{figure*}[ht!]
\centering
\includegraphics[scale=0.66]{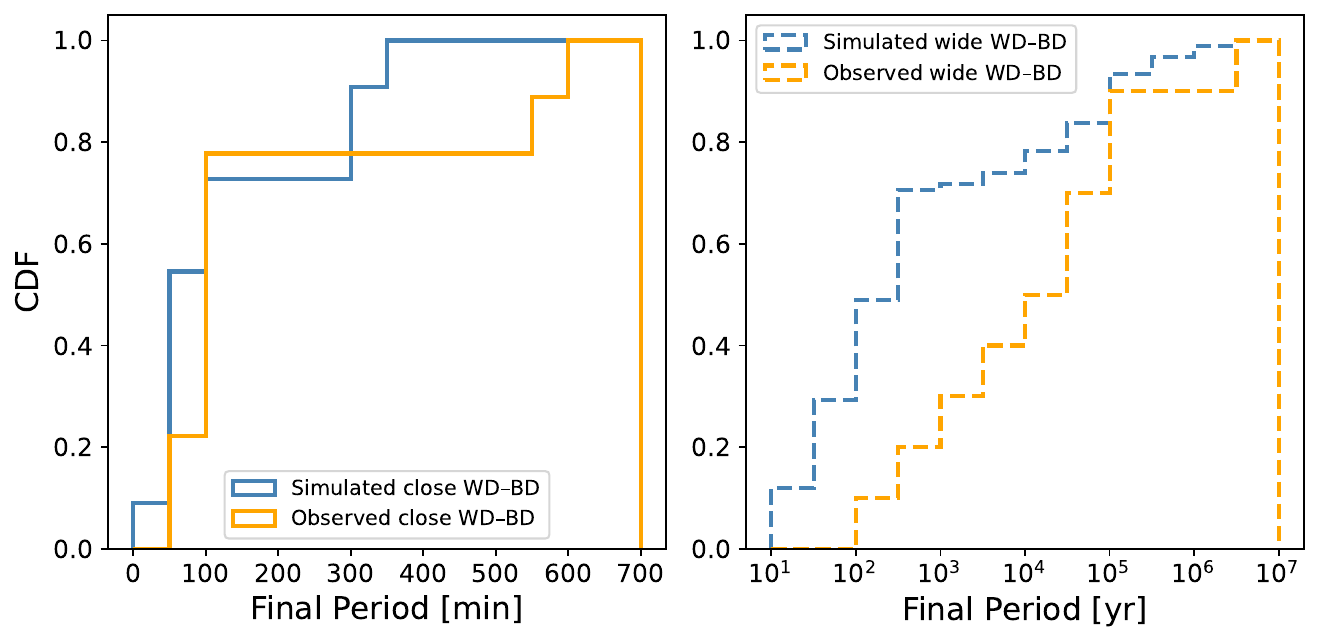}
\caption{
The CDFs of orbital period of WD--BD binaries.
The left panel compares the orbital period distribution of close WD--BD binaries between the simulation and observation samples, 
while the right panel shows the comparison for wide WD--BD samples.
}
\label{fig: period_comparison}
\end{figure*}

To further examine the binary evolutionary models, we compare the orbital period distributions between the simulated and observed WD--BD binaries.
Figure~\ref{fig: period_comparison} presents the CDFs of orbital periods of close and wide systems in left and right panels, respectively. The blue and orange lines in each panel represent the simulated and observed systems.
We find that for the close systems, the CDFs in simulation and observation match well, with a Kolmogorov-Smirnov test p-value of 0.59, indicating that the adopted CE model and $\alpha_{\rm CE}$ value are appropriate.
However, for wide systems, the simulated and observed samples show a discrepancy in the CDFs of period. Since the wide binaries are difficult to be detected, this difference is likely attributed to the observational biases from different detection methods, as further discussed in Section~\ref{sec:disc3}.

\section{Discussion} \label{sec:disc}

In this section, we discuss the implications of our population synthesis results for the formation and observability of WD--BD binaries. Our models predict that the orbital period distribution is primarily shaped by whether systems undergo CE evolution, leading to distinct post-CE and no-CE populations. This naturally gives rise to a deficit of systems at intermediate periods, whose interpretation requires careful consideration in light of current observational limitations. We therefore assess the robustness of this predicted period structure, examine the main sources of model uncertainty, and compare our results with previous theoretical and observational studies. We conclude by discussing how future observations can test these predictions and refine our understanding of the formation and evolution of WD--substellar binaries.

\subsection{Period Gap} \label{sec:disc-1}

Our comprehensive investigation into the post-MS evolution of observed MS--BD binaries has yielded crucial insights into the formation and ultimate fates of substellar companions around WDs. 
Based on an MS--BD binaries catalog and theoretical evolutionary model, we predict the future period-mass distribution of WD--BD binaries and identify a period gap \citep[see also][]{Nordhaus13}, which is broadly consistent with current observations.

However, the apparent ``period gap'' around 1--1000~days in the distribution of WD--BD binaries should be interpreted with appropriate caution. In our evolutionary framework, a deficit of systems with orbital periods of $\sim$1--1000~days arises naturally from binary evolution physics: systems that undergo CE evolution are driven to very short periods, whereas systems that avoid CE evolution experience orbital expansion due to post-MS mass loss, resulting in much wider separations. Consequently, WD--BD binaries at intermediate periods are intrinsically rare in the simulations. The currently known WD--BD systems are broadly consistent with this theoretical expectation, with observed objects located on either side of the predicted gap. However, the total number of confirmed WD--BD binaries remains small, and observational selection effects cannot be excluded. Radial velocity and transit surveys preferentially detect short-period systems, while direct imaging and astrometric methods are most sensitive to wide companions, potentially producing an apparent deficit at intermediate periods.

A noteworthy exception is the transiting planetary-mass companion WD~1856+534~b \citep{Vanderburg20, Lagos2021}. With an orbital period of $\sim$1.4~days, it resides near the boundary of the predicted gap. We emphasize, however, that WD~1856+534~b has a mass well below the BD regime and is therefore not directly comparable to the WD--BD population considered in this work. Owing to its lower mass, the minimum survivable post-CE orbital period for such a planet is longer than that for typical BDs. Moreover, its formation has been suggested to involve more complex evolutionary pathways, such as dynamically assisted or multi-body CE evolution \citep[e.g.,][]{Chamandy21}. As such, WD~1856+534~b does not contradict the predicted WD--BD period gap, but instead highlights the diversity of evolutionary outcomes for substellar companions. Overall, while current observations do not contradict the existence of a WD--BD period gap, larger and more complete samples will be required to determine whether it represents an intrinsic outcome of binary evolution or is largely shaped by observational biases.

\subsection{Model Uncertainties} \label{sec:disc0}
While our current model provides a robust framework for predicting the evolution of MS--BD binaries, it is important to acknowledge areas where further refinement can enhance the fidelity of our predictions.
A primary assumption in our treatment of orbital evolution during stellar mass loss is the adiabatic $Jeans$ mode approximation \citep{Huang63}. This approximation, valid when the mass-loss timescale significantly exceeds the orbital period, maintains a constant eccentricity and precludes companion ejection. While appropriate for close binaries, this assumption may break down for BDs in wide, particularly highly eccentric, orbits around evolving stars. In such scenarios, the stellar mass-loss timescale can become comparable to the orbital period, leading to non-adiabatic effects that could result in orbital eccentricity excitation or even ejection of the companion, forming free-floating BDs. Future work will explore the parameter space where non-adiabatic effects become dominant, potentially requiring more sophisticated N-body simulations coupled with stellar evolution models to accurately capture these complex dynamical outcomes.

Furthermore, the presence of tertiary or higher-order companions in some MS--BD systems introduces additional dynamical complexities. For simplicity, our current study assumes distant tertiary companions have a negligible gravitational influence on the primary MS--BD binary's evolution. However, multi-body interactions can induce Kozai-Lidov cycles, secular perturbations, or even scattering events that dramatically alter the BD's orbit, particularly during phases of significant stellar mass loss. Incorporating a comprehensive N-body treatment alongside stellar evolution within a larger population synthesis framework will be a critical next step to accurately model these intricate systems and assess the impact of higher-order multiplicity on the ultimate fate of BD companions.

\subsection{Comparison with previous studies} \label{sec:disc1}
Our findings build upon and complement previous efforts to characterize WD--BD binary systems and their progenitors. Earlier studies by \cite{Zorotovic22} and \cite{ChenZL24} leveraged observed WD--BD data to constrain CE parameters and to test hypotheses related to the brown dwarf desert by retro-calculating progenitor systems. In contrast, our approach forward-evolves observed MS--BD systems, offering a predictive framework that allows for direct comparison with existing and future WD--BD observations, thereby validating evolutionary models and anticipating undiscovered populations.

A significant enhancement in our study is the explicit inclusion of tidal effects throughout the primary's post-MS evolution. As demonstrated in Figure~\ref{fig:fate_distribution}, tidal dissipation can induce significant orbital decay and circularization *prior* to the CE phase, influencing whether a system enters CE and its subsequent outcome. Our results indicate that tidal effects lead to an approximately 10\% increase in systems entering the CE phase compared to models neglecting tides. The strength of tidal interactions is highly sensitive to binary separation, scaling inversely with the $7^{\rm th}$ to $8^{\rm th}$ power, meaning their influence is negligible for extremely wide binaries but crucial for closer systems. Given the ongoing debate surrounding tidal models in binary evolution, the increasing number of observed close WD--BD binaries will provide invaluable empirical constraints, allowing us to refine these critical parameters in future simulations.

BDs represent a critical intermediate mass regime, exhibiting evolutionary pathways distinct from lower-mass giant planets. While giant planets are less likely to survive the CE phase due to their lower masses \citep{Nordhaus13}, leading to the observed scarcity of close WD+planet systems \citep{Vanderburg20, Blackman21}, brown dwarfs possess enhanced survival capabilities. The period gap we observe in the WD--BD population, spanning approximately 1 to 1000 days, aligns well with theoretical predictions for substellar companions, with its inner edge exhibiting consistency with findings for giant planets presented by \cite{Nordhaus13}. This result demonstrates the unified physics governing post-CE binary evolution across a range of companion masses.

\subsection{The impact of CE efficiency $\alpha$} \label{sec:disc2}

The CE efficiency parameter, $\alpha_{\rm CE}$, remains a cornerstone in modeling the outcomes of binary evolution involving CE phases. This parameter, which quantifies the efficiency of orbital energy conversion into envelope ejection, critically determines the fate of systems that enter CE. Previous studies, often focused on post-CE binaries containing white dwarfs, suggest a relatively low $\alpha_{\rm CE}$ in the range of 0.2 to 0.4 \citep{Zorotovic10,Toonen13,Camacho14,Parsons15,Hernandez21,Zorotovic22,Scherbak23,ChenZL24}.
Recently, \citet{Ge22} provided a more self-consistent method for calculating the binding energy, which addresses both the remnant core response and the companion mass effect. \citet{Ge24a} further found that the average CE efficiency parameter is 0.32. However, it need not be constant, but a function of the initial mass ratio, based on well-constrained progenitor mass and evolutionary stage.

\begin{figure*}[ht!]
\centering
\includegraphics[scale=0.55]{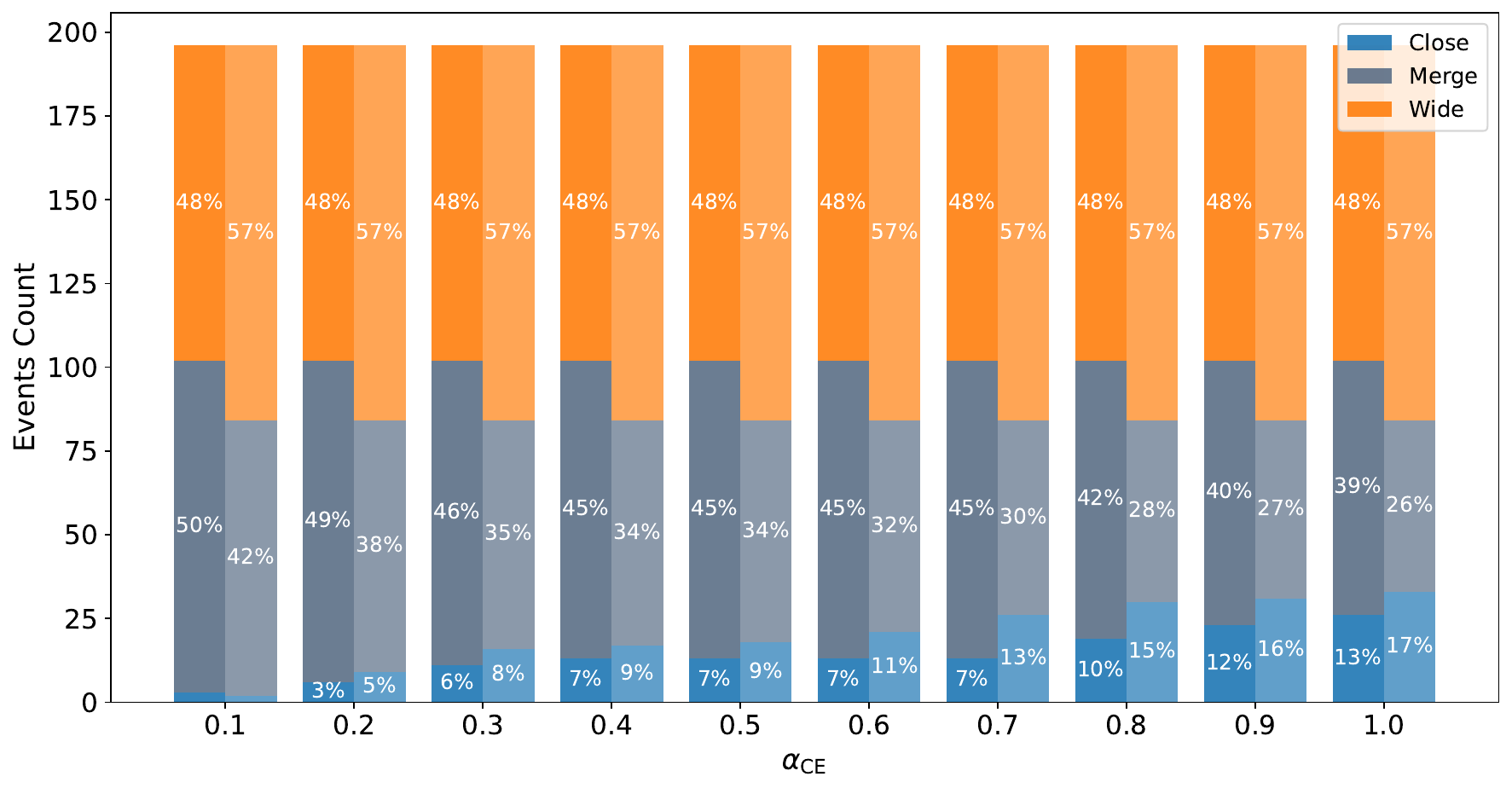}
\caption{
Distribution of the number of known MS--BD binary systems with different evolution outcomes as a function of the CE efficiency parameter $\alpha_{\rm CE}$. The bars represent the counts of three different evolutionary outcomes -  wide WD--BD binaries (orange), merged systems (gray), and detached close WD--BD binaries (blue). For each fixed $\alpha_{\rm CE}$ value, the darker bars show the results with tidal effects, while the lighter bars present the cases without tidal effects. 
}
\label{fig:counts}
\end{figure*}

Our simulations, spanning a range of $\alpha_{\rm CE}$ from 0.1 to 1.0 (Figure~\ref{fig:counts}), vividly illustrate its profound impact. As expected, higher $\alpha_{\rm CE}$ values, indicating more efficient energy transfer, lead to a greater proportion of systems successfully ejecting the CE and surviving as detached close WD--BD binaries. Conversely, lower $\alpha_{\rm CE}$ values result in more systems merging within the envelope or forming extremely close systems that rapidly evolve into CVs. The interplay between tidal dissipation and $\alpha_{\rm CE}$ is also evident: tidal effects, by bringing systems closer, increase the number of binaries entering the CE phase, and thereby amplify the importance of $\alpha_{\rm CE}$ in determining the ultimate distribution of WD--BD systems. Our results highlight the continued necessity of empirically constraining $\alpha_{\rm CE}$ through precise observations of post-CE binaries.

\subsection{Implication for future observations} \label{sec:disc3}
Our study offers guidance for future observational campaigns targeting substellar companions to white dwarfs.
One prediction from our simulations is the existence of a population of WD--BD binaries with orbital periods around 100~years, corresponding to separations of approximately 15~AU. This population remains largely undetected, likely due to observational biases: direct imaging techniques favor very wide separations, while radial velocity and photometric (eclipsing/transiting) methods are most sensitive to short-period systems. However, next-generation high-contrast imaging facilities and precision astrometry missions may be ideally suited to uncover these systems and measure their orbital properties. 
Recently, the imaging observations of nearby metal-polluted WDs with the Mid-Infrared Instrument (MIRI) on the James Webb Space Telescope (JWST) have revealed candidate planetary-mass companions, including Jovian planets, and low-mass BDs \citep{Mullally24, Mullally25, Debes25, Albert25}.
For example, the MIRI Excesses around Degenerates (MEAD) survey of JWST has reported a directly imaged candidate BD companion to a WD, MEAD~62~B \citep{Albert25}. If confirmed, this BD candidate, with a mass of $\sim14~M_{\rm Jup}$ and a projected separation of $\sim40~\rm AU$, would represent a valuable wide WD--BD system.
Furthermore, microlensing surveys, which have already proven successful in detecting BDs around MS stars at these intermediate separations, hold significant promise. The recent discovery of a BD in the KMT-2020-BLG-0104 multi-body system around a WD via microlensing \citep{Zhang24NatAs} validates this pathway, suggesting that systematic microlensing campaigns could bridge this observational gap.

Moreover, our simulations predict that approximately 20 systems will evolve into semi-detached WD--BD binaries, destined to become cataclysmic variables with BD donors (BD-CVs). While a few such systems have been identified \citep{Hernandez16,Harrison16,Pala19,Neustroev23}, their rarity underscores the need for targeted observational efforts. These systems are laboratories for understanding mass transfer from substellar objects to WDs, the evolution of accretion disks, and the physics of highly irradiated BDs. Continued spectroscopic and photometric monitoring with facilities like the HST and the Very Large Telescope (VLT), coupled with efforts to identify new BD-CV candidates from wide-field surveys, will provide invaluable data to test our evolutionary models and refine our understanding of the physical properties of giant planets and brown dwarfs under extreme conditions.

\section{Conclusion} \label{sec:conc}
In this study, Project FOSSO I, we have investigated the long-term orbital evolution of a sample of approximately 200 observed BDs orbiting solar-type MS stars. By employing numerical simulations with the COMPAS code, we traced the evolution of these systems as their host stars transitioned through post-MS phases and ultimately became WDs. Our primary goal was to predict the fates and orbital characteristics of the resulting WD--BD binaries and to compare these predictions with the current observational data.

Our simulations reveal that the initial properties of MS--BD binaries fundamentally dictate their post-MS evolution. We found that these systems generally evolve into either very close or wide WD--BD binaries, forming a period gap between 1 to 1000~days in the distribution of WD--BD binaries. This predicted gap is in agreement with the period distribution of currently observed WD--BD systems, providing strong validation for our evolutionary models.

For close MS--BD systems, the CE phase plays a critical role. Depending on the efficiency of energy transfer ($\alpha_{\rm CE}$), a significant fraction of these systems either merge during the CE phase or emerge as extremely close post-CE binaries. 
Following the work from \citet{Rappaport21}, we identified a minimum orbital period boundary for detached WD--BD systems, 
estimated to be around 120--40 minutes, decreasing with increasing BD mass. 
Systems with periods below this threshold are predicted to evolve into semi-detached systems, potentially forming CVs with BD donors. Our analysis suggests that approximately 20 systems from our initial sample are likely candidates to evolve into such CVs.

Conversely, wide MS--BD systems avoid the CE phase, experiencing orbital expansion due to stellar mass loss during the host star's giant phases. Our models predict a substantial population of WD--BD binaries with orbital periods around 100 years (separations of $\sim15$ AU) that remain largely undiscovered. The current lack of observed systems in this regime is likely attributable to observational biases, as direct imaging typically targets much wider separations, while radial velocity and eclipsing methods favor much shorter periods. Future advancements in high-contrast imaging, astrometry surveys, and microlensing observations hold significant promise for detecting these predicted wide WD--BD systems, potentially expanding the known population. The recent microlensing detection of a BD companion in the KMT-2020-BLG-0104 multi-body system around a WD \citep{Zhang24NatAs} exemplifies the potential of this method.

This study not only enhances our understanding of the long-term evolution of substellar companions but also provides valuable insights for constraining binary evolution parameters, particularly those governing CE physics and tidal interactions. Continued observations of both detached WD--BD binaries and CVs with BD donors will be crucial for further refining these models and for elucidating the full spectrum of fates for brown dwarfs in binary systems, thereby advancing the fields of substellar objects and binary stellar evolution.

\begin{acknowledgements}
      We acknowledge the financial support from the National Key R\&D Program of China (2020YFC2201400), NSFC grant 12073092, 12103097, 12103098, the science research grants from the China Manned Space Project (No. CMS-CSST-2021-B09). 
      HG acknowledges the support from the Strategic Priority Research Program of the Chinese Academy of Sciences (grant No. XDB1160201), the National Natural Science Foundation of China (NSFC Nos. 12288102, 12525304), the National Key R\&D Program of China (No. 2021YFA1600403).
\end{acknowledgements}

   \bibliographystyle{aasjournal}
   \bibliography{aanda}

\begin{appendix}

\onecolumn

\section{MS--BD catalog} \label{apx:MSBD_catalog}
In this appendix, we summarize the properties of the 196 selected MS--BD systems in our sample. 17 of the 213 compiled systems are expected to remain on the main sequence for longer than 13.7~Gyr, and therefore are excluded these systems from our analysis.

\small
\begin{longtable}{lccccccc}
\multicolumn{8}{l}{\textbf{Systems with real-mass measured.} Please note the differences in the units of orbital periods ($P_{\rm orb}$) used in the table (day and yr, respectively).} \\
\toprule
Host star & $M_{\rm BD}$ & $P_{\rm orb}$ & $a$ & $e$ & $M_*$ & [Fe/H] & Ref. \\
          & [$\rm M_{Jupiter}$] & [day]  & [AU] &  & [$\rm M_{\odot}$] & [dex] & \\
\midrule
$^*$BD-14 3065 A & $12.37\pm0.9$ & $4.289\pm5.20\,\mathrm{E}{-6}$ & $0.0656\pm0.003$ & $0.07\pm0.01$ & $1.41$ & $-0.34$ & 1\\
$^*$TIC 4672985 & $12.74\pm1$ & $69.048\pm0.0005$ & $0.33\pm0.02$ & $0.018\pm0.004$ & $1.01$ & $0.14$ & 2\\
$^*$TOI-4603 & $12.89\pm0.58$ & $7.246\pm0.00022$ & $0.0888\pm0.001$ & $0.325\pm0.02$ & $1.752$ & $-0.236$ & 3\\
$^*$HATS-70 & $12.9^{+1.8}_{-1.6}$ & $1.8882\pm1.50\,\mathrm{E}{-6}$ & $0.03632^{+0.00074}_{-0.00087}$ & $0.18^{+0}_{-0.18}$ & $1.78$ & $0.04$ & 4\\
HD 39392 & $17.3\pm3.84$ & $394.3\pm1.4$ & $1.08\pm0.03$ & $0.394\pm0.008$ & $1.08$ & $0.32$ & 5\\
gamma Cephei A & $17.539\pm0.745$ & $905.64\pm2.83$ & $2.1459\pm0.0048$ & $0.0724^{+0.0679}_{-0.0575}$ & $1.4$ & $0.18$ & 6\\
$^*$GPX-1 & $19.7\pm1.6$ & $1.7446\pm8.00\,\mathrm{E}{-6}$ & $0.0338\pm0.0003$ & $0$ & $1.68$ & $0.35$ & 7\\
$^*$Kepler-39 & $20.1\pm1.3$ & $21.087\pm3.70\,\mathrm{E}{-5}$ & $0.164\pm0.003$ & $0.112\pm0.057$ & $1.29$ & $0.1$ & 5\\
$^*$CoRoT-3 & $21.66\pm1$ & $4.2568\pm5.00\,\mathrm{E}{-6}$ & $0.057\pm0.003$ & $0$ & $1.36$ & $0.14$ & 5\\
$^*$TOI-5882 & $22.01^{+0.7}_{-0.6}$ & $7.149\pm1.60\,\mathrm{E}{-5}$ & $0.085\pm0.001$ & $0.0339^{+0.008}_{-0.009}$ & $1.334$ & $0.18$ & 8\\
$^*$TOI-1994 & $22.1\pm2.6$ & $4.0337\pm5.10\,\mathrm{E}{-6}$ & $0.0613\pm0.0019$ & $0.0341\pm0.059$ & $1.86$ & $-0.01$ & 9\\
$^*$Kepler-2002 & $23^{+3}_{-2}$ & $19.005\pm5.00\,\mathrm{E}{-5}$ & $0.1357$ & $0.36\pm0.03$ & $0.91$ & $-0.37$ & 10\\
HD 141937  & $23.5^{+4.7}_{-5.1}$ & $653.5\pm0.3$ & $1.52$ & $0.41\pm0.01$ & $1.1$ & $0.11$ & 11, 5\\
$^*$KELT-1 A & $27.23^{+0.5}_{-0.48}$ & $1.2175\pm1.50\,\mathrm{E}{-5}$ & $0.02472\pm0.00039$ & $0.01^{+0.01}_{-0.007}$ & $1.335$ & $0.052$ & 12\\
$^*$TOI-5422 & $27.7^{+1.4}_{-1.1}$ & $5.3772\pm1.00\,\mathrm{E}{-5}$ & $0.0616^{+0.0016}_{-0.0012}$ & $0.0942^{+0.005}_{-0.004}$ & $1.051$ & $-0.007$ & 13\\
HIP 84056  & $31.9^{+8.5}_{-5.3}$ & $825.83^{+9.8618}_{-8.766}$ & - & $0.079^{+0.06}_{-0.053}$ & $1.69$ & $0.08$ & 5\\
$^*$TOI-4776 A & $32\pm1.9$ & $10.414\pm1.40\,\mathrm{E}{-5}$ & $0.0962\pm0.002$ & $0.014^{+0.02}_{-0.01}$ & $1.06$ & $-0.05$ & 13\\
HD 68638 A & $35.1\pm1.4$ & $240.75\pm0.4$ & $0.757$ & $0.56^{+0.035}_{-0.038}$ & $1$ & $-0.33$ & 14\\
$^*$EPIC 219388192 A & $36\pm1.6$ & $5.2926\pm2.60\,\mathrm{E}{-5}$ & $0.0593\pm0.0013$ & $0.1893\pm0.0021$ & $0.99$ & $0.03$ & 15\\
$^*$Kepler-494 & $37^{+4}_{-2}$ & $8.0251\pm2.27\,\mathrm{E}{-6}$ & $0.08^{+0.03}_{-0.02}$ & $0.26\pm0.02$ & $1.1$ & $0.16$ & 10\\
$^*$WASP-128 & $37.19^{+0.83}_{-0.85}$ & $2.2088\pm4.50\,\mathrm{E}{-7}$ & $0.0359^{+0.00038}_{-0.0004}$ & $0^{+0.007}_{-0}$ & $1.155$ & $0.01$ & 16\\
$^*$Kepler-492 & $39.9\pm1$ & $11.72\pm2.10\,\mathrm{E}{-6}$ & $0.0987$ & $0.0155$ & $0.925$ & - & 17\\
HD 30246  & $42.18\pm0.23$ & $990.08\pm0.58$ & $2.009^{+0.081}_{-0.088}$ & $0.6605\pm0.003$ & $1.03$ & $0.08$ & 18, 5\\
HD 91669  & $43.2\pm2.2$ & $487.5\pm0.6$ & - & $0.448\pm0.002$ & $0.914$ & $0.31$ & 5\\
HIP 117179  & $44.197^{+4.786}_{-5.052}$ & $247.98\pm1.7$ & $0.788$ & $0.416\pm0.077$ & $1.019$ & - & 5, 19\\
$^*$TOI-1406 & $46^{+2.6}_{-2.7}$ & $10.574\pm0.0006$ & $0.101^{+0.0022}_{-0.0026}$ & $0.026^{+0.013}_{-0.01}$ & $1.18$ & $-0.08$ & 20\\
TYC 8321-266-1  & $46.226^{+8.842}_{-8.644}$ & $413.89\pm5.94$ & $1.157$ & $0.505\pm0.077$ & $1.16$ & - & 5\\
$^*$TOI-3755 & $47.1\pm2$ & $5.5437\pm6.20\,\mathrm{E}{-6}$ & $0.0629\pm0.0013$ & $0.005^{+0.0031}_{-0.0026}$ & $1.037$ & $0.334$ & 8\\
ISOY J0535-0447  & $50$ & $3.9056\pm3.00\,\mathrm{E}{-5}$ & $0.0465$ & - & $1.16$ & - & 21\\
$^*$EPIC 212036875 & $52.3\pm1.9$ & $5.1699\pm2.70\,\mathrm{E}{-5}$ & $0.0645\pm0.0011$ & $0.1323^{+0.0042}_{-0.0041}$ & $1.23$ & $-0.303$ & 22, 23\\
$^*$TOI-503 & $53.6\pm1.1$ & $3.6775\pm0.0002$ & - & $0$ & $1.8$ & $0.61$ & 24\\
$^*$TOI-852 & $53.7^{+1.4}_{-1.3}$ & $4.9456\pm8.00\,\mathrm{E}{-5}$ & $0.063$ & $0.0036$ & $1.32$ & $0.334$ & 25\\
$^*$TOI-3577 & $53.8^{+1.9}_{-2.2}$ & $5.2668\pm1.30\,\mathrm{E}{-5}$ & $0.0623\pm0.001$ & $0.006^{+0.008}_{-0.004}$ & $1.11$ & $-0.03$ & 8\\
HD 5433  & $53.8\pm1.7$ & $576.6\pm1.6$ & - & $0.81\pm0.02$ & $0.98$ & - & 5\\
$^*$TOI-2844 & $54^{+4.9}_{-5.1}$ & $3.5524\pm3.00\,\mathrm{E}{-6}$ & $0.0537\pm0.00082$ & $0.42^{+0.046}_{-0.041}$ & $1.585$ & $0.06$ & 8\\
$^*$TOI-811 A & $59.9\pm3.2$ & $25.166\pm4.00\,\mathrm{E}{-5}$ & $0.1874$ & $0.509$ & $1.32$ & $0.4$ & 25\\
$^*$WASP-30 & $60.86\pm0.89$ & $4.1567\pm1.30\,\mathrm{E}{-5}$ & $0.05325\pm0.00039$ & $0$ & $1.166$ & $-0.03$ & 26\\
$^*$KOI-415 & $62.14\pm2.69$ & $166.79\pm0.00022$ & $0.593\pm0.013$ & $0.698\pm0.002$ & $0.94$ & $-0.24$ & 27\\
TYC 3873-761-1  & $63.302^{+8.897}_{-9.128}$ & $526.19\pm3.26$ & $1.255$ & $0.546\pm0.042$ & $0.925$ & - & 5\\
$^*$CoRoT-15 & $63.4^{+4.4}_{-4.2}$ & $3.0604\pm3.00\,\mathrm{E}{-5}$ & $0.045^{+0.014}_{-0.01}$ & $0$ & $1.32$ & $0.1$ & 28, 5\\
$^*$TOI-569 & $64.1\pm1.9$ & $6.556\pm0.00016$ & - & $0^{+0.0035}_{-0}$ & $1.21$ & $0.29$ & 5, 20\\
HD 115517  & $64.475^{+12.737}_{-12.563}$ & $439.49\pm4.94$ & $1.3347$ & $0.403\pm0.061$ & $1$ & - & 5\\
$^*$TOI-4737 & $66.3^{+2.7}_{-3.1}$ & $9.32$ & $0.097$ & $0.0063^{+0.007}_{-0.0042}$ & $1.336$ & $0.24$ & 8\\
HD 156312 B & $66.519^{+10.539}_{-10.74}$ & $238.43\pm1$ & $0.744$ & $0.239\pm0.064$ & $0.99$ & - & 5\\
HD 92320  & $70\pm3.1$ & $145.4\pm0.01$ & - & $0.323\pm0.001$ & $0.92$ & $-0.1$ & 5\\
TYC 7572-327-1  & $70.921^{+8.612}_{-8.963}$ & $406.5\pm4.23$ & - & $0.476\pm0.054$ & $0.9385$ & - & 5\\
HD 132032  & $72.23\pm4.19$ & $274.33\pm0.24$ & - & $0.0844\pm0.0024$ & $1.12$ & $0.22$ & 5\\
$^*$TOI-2533 & $74.9\pm5.3$ & $6.6858\pm1.30\,\mathrm{E}{-5}$ & - & $0.2476$ & $1.02$ & $-0.3$ & 29\\
HD 140913  & $75.36\pm5.15$ & $147.97$ & - & $0.54$ & $0.98$ & $0.07$ & 5\\
\midrule
Host star & $M_{\rm BD}$ & $P_{\rm orb}$ & $a$ & $e$ & $M_*$ & [Fe/H] & Ref. \\
          & [$\rm M_{Jupiter}$] & [yr]  & [AU] &  & [$\rm M_{\odot}$] & [dex] & \\
\midrule
HD 149806 A & $12.597^{+2.694}_{-2.782}$ & $45.233^{+17.594}_{-10.121}$ & $12.404^{+3.103}_{-2.056}$ & $0.357^{+0.162}_{-0.236}$ & $0.93$ & $0.22$ & 18\\
pi Men  & $12.6\pm2$ & $5.7175\pm0.00093087$ & $3.308\pm0.039$ & $0.645\pm0.001$ & $1.094$ & $0.08$ & 30\\
HD 128717  & $12.8$ & $2.9821$ & $2.18$ & - & $1.17$ & - & 19, 31\\
GJ 2030 A & $12.934\pm2.36$ & $70.185\pm6.8446$ & $16.752\pm1.31$ & $0.045\pm0.038$ & $0.96$ & $-0.4$ & 18\\
HD 93351  & $13.07^{+1.993}_{-1.361}$ & $38.358^{+16.242}_{-4.2127}$ & $11.154^{+2.956}_{-0.949}$ & $0.109^{+0.046}_{-0.041}$ & $0.94$ & $-0.23$ & 18\\
HR 2562  & $13.2\pm11$ & - & $22.2^{+3.8}_{-2.9}$ & $0.34^{+0.23}_{-0.18}$ & $1.368$ & $0.1$ & 32, 33\\
HIP 67537  & $13.7^{+5.6}_{-2.4}$ & $6.98^{+0.52}_{-0.29}$ & - & $0.589^{+0.035}_{-0.03}$ & $2.41$ & $0.15$ & 5\\
HD 217786 A & $13.852^{+1.267}_{-1.311}$ & $3.563^{+0.0032854}_{-0.0013689}$ & $2.446^{+0.109}_{-0.119}$ & $0.311^{+0.002}_{-0.003}$ & $1.02$ & $-0.135$ & 18\\
TYC 8998-760-1  & $14\pm3$ & - & $146^{+16}_{-10}$ & $0.44^{+0.17}_{-0.18}$ & $1$ & - & 34\\
HD 23596  & $14$ & $4.2029^{+0.021355}_{-0.025462}$ & $2.694^{+0.107}_{-0.118}$ & $0.282^{+0.017}_{-0.014}$ & $1.27$ & $0.32$ & 35\\
$^*$TOI-201 & $14.2\pm2$ & $7.666^{+0.98563}_{-0.57495}$ & $4.28^{+0.36}_{-0.2}$ & $0.64\pm0.02$ & $1.316$ & $0.24$ & 36\\
HD 14067  & $14.9^{+6.4}_{-4.8}$ & $7.83^{+1.2014}_{-0.91}$ & - & $0.694^{+0.048}_{-0.053}$ & $2.4$ & $-0.1$ & 5\\
BD+60 1417  & $15\pm5$ & - & $1662$ & - & $1$ & - & 37\\
HD 68988  & $15.03^{+0.9}_{-0.75}$ & $38.074^{+5.2964}_{-4.8966}$ & $12.1\pm1.1$ & $0.399^{+0.046}_{-0.049}$ & $1.2$ & $0.24$ & 38\\
HD 204313 e & $15.317^{+4.89}_{-5.183}$ & $20.056^{+1.0949}_{-1.0105}$ & $7.457^{+0.399}_{-0.427}$ & $0.253^{+0.071}_{-0.065}$ & $1.045$ & $0.18$ & 18\\
HD 33636  & $15.4$ & $7.7426$ & $3.27\pm0.19$ & $0.55$ & $1.09$ & $-0.06$ & 35\\
HD 73256  & $16\pm1$ & $7.3648^{+0.16427}_{-0.27926}$ & $3.8$ & $0.16\pm0.07$ & $1.05$ & $0.29$ & 39\\
HD 25912  & $16.527^{+1.376}_{-1.399}$ & $72.029^{+5.7363}_{-7.6512}$ & $17.848^{+1.147}_{-1.484}$ & $0.051^{+0.048}_{-0.025}$ & $1.08$ & $0.12$ & 18\\
HIP 99770  & $17\pm6$ & - & $15.8^{+1.7}_{-1}$ & $0.31\pm0.12$ & $1.9$ & - & 40\\
kappa And  & $17.3\pm1.8$ & $416.71^{+149.9}_{-209.86}$ & $78.9^{+18}_{-28}$ & $0.74^{+0.098}_{-0.075}$ & $2.768$ & $-0.36$ & 41\\
HIP 54515  & $17.7^{+7.6}_{-4.9}$ & $88.999^{+41.068}_{-24.093}$ & $24.8^{+7.2}_{-4.7}$ & $0.42\pm0.14$ & $1.88$ & - & 42\\
HD 62364  & $17.71^{+0.85}_{-0.83}$ & $14.12^{+0.069952}_{-0.063956}$ & $6.41^{+0.1}_{-0.11}$ & $0.587^{+0.017}_{-0.015}$ & $1.157$ & $-0.11$ & 38\\
HD 202206 (AB)  & $17.9^{+2.9}_{-1.8}$ & $3.4497\pm0.030116$ & $2.41$ & $0.22\pm0.03$ & $1.13$ & $0.37$ & 43\\
HD 139357  & $18.2^{+6.2}_{-5.1}$ & $3.088\pm0.024$ & - & $0.102\pm0.021$ & $1.35$ & $-0.13$ & 5\\
HD 214823 A & $18.61^{+4.14}_{-1.07}$ & $5.0783^{+0.0037153}_{-0.0038795}$ & $3.08^{+0.126}_{-0.139}$ & $0.161\pm0.003$ & $1.31$ & $0.17$ & 18\\
HD 165131  & $19^{+2}_{-1}$ & $6.4148$ & $3.59$ & $0.67$ & $1.06$ & $0.06$ & 39\\
HD 72659  & $19.4^{+0.8}_{-0.5}$ & $97.034^{+3.3977}_{-2.4983}$ & $21.5^{+0.5}_{-0.4}$ & $0.114^{+0.002}_{-0.003}$ & $1.033$ & $-0.02$ & 44\\
HIP 79098 (AB) & $20.5\pm4.5$ & - & $345\pm6$ & - & $2.636$ & $0.033$ & 45\\
HIP 77900  & $21\pm1$ & - & $3319\pm28$ & - & $3.74$ & - & 46\\
HD 97037  & $21.105^{+1.96}_{-2.07}$ & $17.657^{+0.38877}_{-0.42163}$ & $6.925^{+0.283}_{-0.318}$ & $0.357^{+0.027}_{-0.03}$ & $1.06$ & $-0.077$ & 18\\
HD 66428  & $21.6^{+13.6}_{-8.2}$ & $70.951^{+68.953}_{-28.98}$ & $18^{+10}_{-5.4}$ & $0.24^{+0.17}_{-0.1}$ & $1.14552$ & $0.31$ & 38\\
HIP 73990 A & $22^{+35}_{-6}$ & - & $32\pm7$ & - & $1.72$ & - & 47\\
Kepler-448  & $22^{+7}_{-5}$ & $6.8446^{+6.5708}_{-1.9165}$ & $4.2^{+2.4}_{-0.9}$ & $0.65^{+0.13}_{-0.09}$ & $1.454$ & $0.08146$ & 48\\
HIP 97233  & $22.6^{+5.7}_{-3.3}$ & $2.886^{+0.029}_{-0.024}$ & - & $0.633^{+0.055}_{-0.042}$ & $1.74$ & $0.29$ & 5\\
HD 221420  & $22.9\pm2.2$ & $27.619^{+2.4504}_{-1.5387}$ & $10.15$ & $0.14\pm0.04$ & $1.67$ & $0.29$ & 49\\
ups And A & $23.58^{+2.93}_{-2.29}$ & $3.5084\pm0.0054757$ & $2.55$ & $0.316^{+0.006}_{-0.07}$ & $1.27$ & $0.09$ & 50\\
HIP 39017  & $23.6^{+9.1}_{-7.4}$ & $84.873^{+52.019}_{-26.01}$ & $22.1^{+8}_{-4.9}$ & $0.55^{+0.35}_{-0.4}$ & $1.52$ & - & 51\\
HD 80913  & $23.614^{+3.594}_{-3.861}$ & $29.678^{+7.2841}_{-5.4023}$ & $9.95^{+1.683}_{-1.41}$ & $0.313^{+0.152}_{-0.131}$ & $1.11$ & $-0.68$ & 18\\
HD 206893  & $23.88^{+0.86}_{-2.47}$ & $25.599\pm1.1992$ & $8.93^{+1.4}_{-0.19}$ & $0.27^{+0.04}_{-0.23}$ & $1.24$ & - & 52\\
HD 126053  & $23.9^{+3.1}_{-4}$ & - & $2630$ & - & $1.02$ & $-0.31$ & 53\\
HIP 21152  & $24^{+6}_{-4}$ & $59.001^{+30.992}_{-16.003}$ & $17^{+5}_{-4}$ & $0.36^{+0.37}_{-0.25}$ & $1.44$ & - & 54\\
CD-25 11378  & $24.8\pm1$ & - & $1855$ & - & $1.135$ & - & 55\\
HD 105618  & $26.491^{+4.34}_{-2.827}$ & $211.13^{+63.068}_{-28.121}$ & $35.731^{+6.866}_{-3.639}$ & $0.126^{+0.098}_{-0.08}$ & $1.01$ & - & 18\\
HD 457  & $26.572^{+18.728}_{-11.952}$ & $55.646^{+31.441}_{-19.272}$ & $15.369^{+5.561}_{-3.797}$ & $0.565^{+0.195}_{-0.245}$ & $1.15$ & $0.34$ & 18\\
PZ Tel  & $27^{+25}_{-9}$ & $119.92^{+109.92}_{-29.979}$ & $27^{+14}_{-4}$ & $0.52^{+0.08}_{-0.1}$ & $1.13$ & $0.05$ & 56\\
HD 110537  & $27.099^{+12.755}_{-8.99}$ & $66.994^{+27.441}_{-21.792}$ & $16.714^{+4.538}_{-3.907}$ & $0.401^{+0.115}_{-0.152}$ & $1.02$ & $0.09$ & 18\\
HD 60584  & $28\pm9$ & - & $16.58\pm0.15$ & - & $1.36$ & - & 57\\
HIP 74865  & $28^{+37}_{-10}$ & - & $23\pm6$ & - & $1.42$ & - & 47\\
HIP 78530  & $28\pm10$ & - & $710\pm60$ & - & $2.5$ & - & 58\\
HD 284149 (AB) & $28.26^{+14}_{-18}$ & - & $431.16$ & - & $1.13$ & - & 59\\
eta Tel  & $29^{+16}_{-13}$ & $2201.2\pm487.34$ & $218^{+180}_{-41}$ & $0.34\pm0.26$ & $2.09$ & - & 60\\
HD 188641  & $29.365^{+5.984}_{-3.712}$ & $39.928^{+9.1121}_{-7.163}$ & $12.027^{+1.897}_{-1.567}$ & $0.034^{+0.059}_{-0.088}$ & $1.07$ & $-0.08$ & 18\\
HD 168443  & $29.463^{+12.184}_{-6.128}$ & $4.7908\pm0.0015606$ & $2.8373\pm0.018$ & $0.2113\pm0.0017$ & $0.995$ & $0.04$ & 61\\
HD 188769 (AB) & $30$ & - & $25000$ & - & $1.53$ & $-1.13$ & 62\\
HT Lup (AB) & $30.8\pm3.6$ & - & $7570$ & - & $1.6$ & - & 55\\
OGLE-2014-BLG-1112L  & $31.7\pm8.2$ & - & $9.63\pm1.33$ & - & $1.07$ & - & 63\\
HD 205158  & $31.894^{+3.657}_{-3.813}$ & $44.089^{+3.8442}_{-3.6151}$ & $13.171^{+0.931}_{-0.997}$ & $0.031^{+0.027}_{-0.019}$ & $1.16$ & - & 18\\
HD 85472  & $31.942^{+9.958}_{-10.039}$ & $17.222^{+1.6953}_{-3.0669}$ & $7.661^{+0.501}_{-0.98}$ & $0.264^{+0.064}_{-0.114}$ & $1.36$ & $-0.02$ & 18\\
HIP 64892  & $33\pm4$ & - & $159\pm12$ & - & $2.35$ & - & 64\\
HD 4113 A & $33.8^{+18.8}_{-13.6}$ & $348.32^{+21.932}_{-15.139}$ & $50.438^{+2.06}_{-1.42}$ & $0.648^{+0.008}_{-0.007}$ & $1$ & $0.235$ & 18\\
HD 125390  & $34^{+3.4}_{-3.3}$ & $4.812\pm0.013$ & - & $0.598^{+0.022}_{-0.017}$ & $1.6$ & $-0.11$ & 5\\
HIP 75056 A & $35\pm10$ & $130.05\pm30.116$ & $22.5\pm1$ & $0.44\pm0.03$ & $2.15$ & - & 65\\
GJ 758  & $35.91^{+0.69}_{-0.65}$ & $142.9^{+46.968}_{-26.982}$ & $27.4^{+5.6}_{-3.6}$ & $0.277^{+0.17}_{-0.091}$ & $0.97$ & $0.18$ & 38\\
HD 122562  & $37.203^{+3.67}_{-2.595}$ & $7.9603^{+0.10021}_{-0.082409}$ & $4.313^{+0.088}_{-0.086}$ & $0.706^{+0.008}_{-0.004}$ & $1.13$ & $0.31$ & 18\\
HD 75881  & $38.255^{+14.558}_{-12.009}$ & $57.231^{+0.048731}_{-0.27473}$ & $22.69^{+8.08}_{-4.194}$ & $0.732^{+0.102}_{-0.139}$ & $1.17$ & $0.07$ & 18\\
zet Del  & $40^{+15}_{-5}$ & - & $907^{+723}_{-236}$ & - & $2.5$ & $-0.05$ & 66\\
USco 1602-2401  & $41^{+20}_{-13}$ & - & $1000\pm140$ & - & $1.34$ & - & 67\\
HD 136118  & $42^{+11}_{-18}$ & $3.3101\pm0.065708$ & $1.45\pm0.25$ & $0.352\pm0.006$ & $1.24$ & $-0.065$ & 68\\
HD 23965  & $43.9^{+2.9}_{-2.7}$ & $12.172^{+1.9986}_{-0.41971}$ & $5.58^{+0.6}_{-0.15}$ & $0.805^{+0.012}_{-0.01}$ & $1.2$ & $0.01$ & 69\\
2MASS J1233-5714  & $46.5\pm3.3$ & - & $1368$ & - & $1$ & - & 55\\
HD 213519 A & $47.214^{+5.901}_{-6.425}$ & $88.775^{+5.0084}_{-3.5775}$ & $20.316^{+1.128}_{-1.034}$ & $0.423^{+0.072}_{-0.05}$ & $1.05$ & $0.01$ & 18\\
HD 97334 C & $47.7\pm1.9$ & $15.639$ & $2.85\pm0.03$ & $0.106$ & $1.07$ & $0.11$ & 70\\
HD 104289  & $49.483^{+6.52}_{-6.615}$ & $3.3767\pm0.44794$ & $2.42$ & $0.377\pm0.0672$ & $1.213$ & - & 5\\
HD 210797  & $49.993^{+9.457}_{-9.417}$ & $63.147^{+15.899}_{-9.053}$ & $15.564^{+2.44}_{-1.751}$ & $0.406^{+0.121}_{-0.079}$ & $0.91$ & - & 18\\
HD 13724  & $50.5\pm3.3$ & $123\pm41.068$ & $26.3^{+2.6}_{-0.9}$ & $0.63\pm0.067$ & $1.14$ & $0.23$ & 18\\
HD 97334  & $51.5^{+1.7}_{-1.8}$ & - & $1970\pm20$ & - & $1.07$ & $0.11$ & 71\\
HD 33632 A & $51.7^{+2.6}_{-2.5}$ & $96.999^{+19.001}_{-20.999}$ & $24.1^{+2}_{-2.8}$ & $0.09^{+0.1}_{-0.062}$ & $1.1$ & $-0.22$ & 72\\
HD 46588  & $52\pm18$ & - & $1420$ & - & $1.09$ & $-0.25$ & 73\\
TYC 8252-533-1  & $52^{+17}_{-11}$ & - & $570$ & - & $1.2$ & - & 55\\
HD 149782  & $53.632^{+6.266}_{-7.801}$ & $19.821^{+2.584}_{-2.7515}$ & $7.406^{+0.685}_{-0.758}$ & $0.04^{+0.028}_{-0.014}$ & $1.02$ & - & 18\\
HIP 65423  & $55\pm10$ & - & $228\pm49$ & - & $1.04$ & - & 74\\
HIP 65517  & $55^{+30}_{-20}$ & - & $39\pm7$ & - & $1.01$ & - & 74\\
HIP 79797 B & $55\pm20$ & - & $350$ & - & $1.76$ & - & 75\\
HD 211847  & $55.32^{+1.335}_{-18.488}$ & $9.4278^{+0.72358}_{-1.3648}$ & - & $0.419^{+0.035}_{-0.064}$ & $0.94$ & $-0.08$ & 5\\
HD 14348  & $55.5^{+2.8}_{-2.7}$ & $13.001\pm0.01399$ & $5.96\pm0.13$ & $0.4574\pm0.0036$ & $1.13$ & - & 76\\
beta Cir  & $56\pm7$ & - & $6656$ & - & $1.96$ & - & 77\\
HD 340935  & $58\pm6$ & $3.2203$ & $2.22$ & - & $1.01$ & - & 19\\
HD 26161 A & $58.65\pm18$ & $95.934^{+123.92}_{-44.969}$ & $23.2$ & $0.868^{+0.079}_{-0.12}$ & $1.3$ & $0$ & 38\\
HN Peg  & $59.6^{+4.8}_{-5.9}$ & - & $795\pm15$ & - & $1.08$ & $-0.036$ & 53\\
HD 130948  & $59.8\pm0.6$ & $35076$ & $1100$ & - & $1.12$ & $0.06$ & 78\\
HD 175679  & $59.9^{+10}_{-8.2}$ & $3.741^{+0.018}_{-0.017}$ & - & $0.3807^{+0.0084}_{-0.0081}$ & $2.7$ & $-0.14$ & 5\\
HD 23514  & $60\pm10$ & $7005.2^{+2498.3}_{-3897.3}$ & $417^{+100}_{-160}$ & $0.29^{+0.11}_{-0.28}$ & $1.211$ & $-0.0388$ & 79\\
HD 40781  & $60$ & - & $3300$ & - & $1.09$ & - & 62\\
HD 51400  & $60$ & - & $2300$ & - & $0.93$ & - & 62\\
TYC 8025-428-1  & $60$ & - & $5000$ & - & $0.9$ & $0.06$ & 62\\
HD 167665  & $60.3\pm0.7$ & $12.131^{+0.0029837}_{-0.0029979}$ & $5.5^{+0.045}_{-0.04}$ & $0.339\pm0.0034$ & $1.054$ & $-0.14$ & 80\\
HD 984  & $61\pm4$ & $139.9^{+49.966}_{-29.979}$ & $28^{+7}_{-4}$ & $0.76\pm0.05$ & $1.2$ & $-0.01$ & 81\\
HIP 89620  & $61.29^{+1.05}_{-0.96}$ & $12.134\pm0.015$ & $5.604\pm0.05$ & $0.3411$ & $1.137$ & - & 38\\
HIP 59173 A & $62\pm27$ & - & $5.5\pm2.4$ & - & $5.23$ & - & 82\\
HD 74014  & $62\pm1.6$ & $18.56^{+0.12}_{-0.11}$ & - & $0.5206^{+0.0021}_{-0.002}$ & $1$ & $0.26$ & 5\\
HD 6718  & $62.79^{+16.98}_{-13.8}$ & $6.8337\pm0.48186$ & $3.56\pm0.15$ & $0.1\pm0.04$ & $0.96$ & $-0.06$ & 68\\
HIP 29724  & $63\pm8$ & - & $6.3\pm0.09$ & - & $1.08$ & - & 57\\
HD 49197  & $63.2^{+12.6}_{-26.32}$ & $145.9^{+52.964}_{-66.954}$ & $29.1^{+6.7}_{-9.7}$ & $0.28^{+0.29}_{-0.2}$ & $1.21$ & - & 83\\
HD 168769  & $63.983^{+10.48}_{-6.315}$ & $25.426^{+1.7859}_{-1.612}$ & $8.399^{+0.518}_{-0.561}$ & $0.181^{+0.097}_{-0.072}$ & $0.91$ & $0.03$ & 18\\
HIP 98819  & $71.4\pm1$ & $71.9^{+2.4}_{-2.2}$ & $17.64\pm0.3$ & $0.5081$ & $0.994$ & - & 38\\
HD 48679  & $71.6^{+7}_{-6.6}$ & $3.0434^{+0.00064}_{-0.00062}$ & - & $0.82467^{+0.00048}_{-0.0005}$ & $1.03$ & $0.21$ & 5\\
HD 19467  & $71.6^{+5.3}_{-4.6}$ & $319^{+114}_{-72}$ & $46.9^{+11}_{-7.4}$ & $0.5$ & $0.96$ & - & 84\\
HD 72946  & $72.4\pm1.6$ & $15.919^{+0.15058}_{-0.12868}$ & - & $0.495\pm0.006$ & $0.96$ & $0.11$ & 5\\
HR 7672  & $72.7\pm0.8$ & $63.028\pm4.7748$ & $16.6\pm2.1$ & $0.468$ & $1.08$ & $0.07$ & 85\\
HD 8291  & $74.6^{+6.1}_{-29}$ & - & $2570$ & - & $1.03$ & $-0.0717$ & 86\\
\midrule
\multicolumn{8}{l}{$^*$ denote the transiting systems}\\
\multicolumn{8}{p{\textwidth}}{\textbf{Reference. }
1.\cite{Subjak24};
2.\cite{Jones24};
3.\cite{Khandelwal23};
4.\cite{Zhou19};
5.\cite{Stevenson23};
6.\cite{Huang22};
7.\cite{Benni21};
8.\cite{Vowell25};
9.\cite{Page24};
10.\cite{Canas23};
11.\cite{Wallace25};
12.\cite{Siverd12};
13.\cite{Zhang25};
14.\cite{Unger23};
15.\cite{Nowak17};
16.\cite{Hodzic18};
17.\cite{Diaz13};
18.\cite{Feng22};
19.\cite{Sahlmann25};
20.\cite{Carmichael20};
21.\cite{Morales12};
22.\cite{Persson19};
23.\cite{Carmichael19};
24.\cite{Subjak20};
25.\cite{Carmichael21};
26.\cite{Anderson11};
27.\cite{Moutou13};
28.\cite{Bouchy11b};
29.\cite{Ferreira24};
30.\cite{Hatzes22};
31.\cite{Pinamonti25};
32.\cite{Roberts25};
33.\cite{Godoy24};
34.\cite{Bohn20};
35.\cite{Kiefer25};
36.\cite{Maciejewski25};
37.\cite{Faherty21};
38.\cite{An25};
39.\cite{Philipot23};
40.\cite{Winterhalder25};
41.\cite{Godoy25};
42.\cite{Currie25};
43.\cite{Benedict17};
44.\cite{Ruggieri25};
45.\cite{Janson19};
46.\cite{Michel24};
47.\cite{Hinkley15};
48.\cite{Masuda17};
49.\cite{Venner21};
50.\cite{Barnes11};
51.\cite{Tobin24};
52.\cite{Milli17};
53.\cite{Baig24};
54.\cite{Kuzuhara22};
55.\cite{Bohn22};
56.\cite{Franson23};
57.\cite{Bonavita22};
58.\cite{Petrus20};
59.\cite{Hoch23};
60.\cite{Chai24};
61.\cite{Pilyavsky11};
62.\cite{Rothermich24};
63.\cite{Han17};
64.\cite{Cheetham18};
65.\cite{Balmer24};
66.\cite{Rosa14};
67.\cite{Aller13};
68.\cite{Kiefer21};
69.\cite{Kiefer19};
70.\cite{Dupuy14};
71.\cite{Dupuy17};
72.\cite{Morsy25};
73.\cite{Subjak23};
74.\cite{Janson12};
75.\cite{Nielsen13};
76.\cite{Xiao23};
77.\cite{Smith15};
78.\cite{Briesemeister19};
79.\cite{Rodriguez12};
80.\cite{Maire24};
81.\cite{Franson22};
82.\cite{Gratton23};
83.\cite{Clarissa23};
84.\cite{Hoch24};
85.\cite{Kasagi25};
86.\cite{Dalal21}.
}
\\
\\
\multicolumn{8}{l}{\textbf{Systems with only $M \sin(i)$ values. Please note the differences in the units of orbital periods ($P_{\rm orb}$) used in the table (day and yr, respectively).}} \\
\midrule
Host star & $M_{\rm BD} \sin(i)$ & $P_{\rm orb}$ & $a$ & $e$ & $M_*$ & [Fe/H] & Ref. \\
          & [$\rm M_{Jupiter}$] & [day]  & [AU] &  & [$\rm M_{\odot}$] & [dex] & \\
\midrule
HD 38801  & $10.13\pm0.23$ & $686.8\pm1.4$ & $1.66\pm0.11$ & $0.059\pm0.026$ & $1.26$ & $0.26$ & 87\\
NGC 2423 3  & $10.6$ & $714.3\pm5.3$ & $2.1$ & $0.21\pm0.07$ & $2.4$ & $0.14$ & 88\\
TYC 4282-605-1  & $10.78\pm0.12$ & $101.54\pm0.05$ & $0.422\pm0.009$ & $0.28\pm0.01$ & $0.97$ & $-0.07$ & 89\\
FN Lyr  & $11.3$ & $794.8\pm15.4$ & $1.88$ & $0.28\pm0.12$ & $1.406$ & $-0.14$ & 90\\
TOI-1444  & $11.8\pm2.9$ & $16.086\pm0.024$ & - & $0$ & $0.934$ & $0.13$ & 91\\
HD 87646 A & $12.4\pm0.7$ & $13.481\pm0.001$ & $0.117\pm0.003$ & $0.05\pm0.02$ & $1.12$ & $-0.17$ & 92\\
MARVELS-17  & $13.1\pm1.6$ & $103.4\pm0.9$ & $0.45\pm0.01$ & $0.24\pm0.08$ & $1.12$ & $0.06$ & 93\\
HAT-P-13  & $14.28\pm0.28$ & $446.27\pm0.22$ & $1.186\pm0.033$ & $0.6616\pm0.0054$ & $1.22$ & $0.43$ & 94\\
MARVELS-10  & $14.8\pm2.2$ & $217.3\pm1.1$ & $0.71\pm0.02$ & $0.53\pm0.15$ & $1.03$ & $-0.43$ & 93\\
del Vir  & $15.83^{+2.3}_{-2.7}$ & $466.63^{+1.5}_{-1.3}$ & $1.33^{+0.08}_{-0.11}$ & $0.36^{+0.06}_{-0.11}$ & $1.4$ & $-0.06$ & 95\\
MARVELS-19  & $17.3\pm2.4$ & $122.2\pm1.7$ & $0.52\pm0.03$ & $0.53\pm0.1$ & $1.33$ & $0.09$ & 93\\
AS 205 A & $19.25\pm1.96$ & $24.84\pm0.03$ & $0.162\pm0.04$ & $0.35\pm0.06$ & $0.9$ & $0$ & 96\\
HD 180314  & $22.707^{+0.252}_{-0.255}$ & $396.03\pm0.62$ & $1.4$ & $0.257\pm0.01$ & $2.6$ & $0.2$ & 97\\
MARVELS-18  & $24.4\pm2.4$ & $117.6\pm1.1$ & $0.51\pm0.02$ & $0.46\pm0.09$ & $1.25$ & $0.11$ & 93\\
HD 137510  & $27.3\pm1.9$ & $801.3\pm0.45$ & $1.88\pm0.064$ & $0.3985\pm0.0073$ & $1.36$ & $0.38$ & 98\\
HD 71827 A & $27.43\pm0.89$ & $15.052\pm5.10\,\mathrm{E}{-5}$ & $0.1228\pm0.002$ & $0.07645\pm0.00068$ & $1.15$ & $-0.11$ & 69\\
MARVELS-6  & $31.7\pm2$ & $47.893^{+0.0063}_{-0.0062}$ & $0.2673$ & $0.1442^{+0.0078}_{-0.0073}$ & $1.11$ & $0.4$ & 99\\
MARVELS-12  & $32.8\pm1.1$ & $252.5\pm0.1$ & $0.85\pm0.03$ & $0.6\pm0.01$ & $1.29$ & $0.36$ & 93\\
HD 119445  & $37.6\pm2.6$ & $410.2\pm0.6$ & $1.71\pm0.06$ & $0.082\pm0.007$ & $3.9$ & $0.04$ & 100\\
MARVELS-8  & $37.7\pm0.4$ & $36.045\pm0.005$ & $0.215\pm0.006$ & $0.101\pm0.004$ & $1.02$ & $0.2$ & 93\\
TYC 2534-698-1  & $39.05\pm11.5$ & $103.7\pm0.111$ & $0.4417$ & $0.385\pm0.011$ & $1.014$ & $-0.14$ & 101\\
MARVELS-4  & $40\pm2.5$ & $9.009\pm0.0004$ & $0.09\pm0.003$ & $0.226\pm0.011$ & $1.16$ & $-0.23$ & 102\\
HD 30774  & $41.59^{+3.3}_{-3.42}$ & $4.93\pm0.01$ & $0.06\pm0.01$ & $0.003\pm0.001$ & $1.04$ & $0.0794$ & 103\\
MARVELS-13  & $41.8\pm2.9$ & $147.6\pm0.3$ & $0.54\pm0.02$ & $0.5\pm0.04$ & $0.96$ & $-0.17$ & 93\\
HD 149790  & $42\pm0.3$ & $6.673\pm0.003$ & - & $0.005\pm0.005$ & $1.45$ & - & 104\\
MARVELS-16  & $42.7\pm0.7$ & $31.656\pm0.007$ & $0.2\pm0.006$ & $0.69\pm0.006$ & $1.06$ & $0.33$ & 93\\
Chang 134  & $43\pm8$ & $82\pm0.3$ & - & - & $1.38$ & - & 105\\
HD 134113  & $47\pm3$ & $201.68\pm0.004$ & $0.638\pm0.01$ & $0.0891\pm0.002$ & $1.177$ & $-1.043$ & 106\\
HD 14651 A & $47\pm3.4$ & $79.418\pm0.0021$ & $0.361\pm0.013$ & $0.4751\pm0.001$ & $0.96$ & $-0.04$ & 107\\
HD 160508  & $48\pm3$ & $178.9\pm0.0074$ & $0.68\pm0.02$ & $0.5967\pm0.0009$ & $1.21$ & $0.01$ & 106\\
HD 16702  & $48.7\pm3.4$ & $72.832\pm0.0023$ & $0.344\pm0.012$ & $0.1373\pm0.0017$ & $1.07$ & $-0.1$ & 107\\
HD 153252 A & $50$ & $5.5262\pm0.000142$ & $0.0652$ & $0.0302\pm0.032$ & $1.162$ & - & 108\\
HIP 65040 A & $50$ & $305.11\pm0.44$ & $0.88$ & $0.591\pm0.059$ & $0.93$ & - & 109\\
MARVELS-11  & $52.1\pm0.6$ & $11.612\pm0.0003$ & $0.1\pm0.003$ & $0\pm0.002$ & $0.99$ & $-0.34$ & 93\\
HD 87646 A & $57\pm3.7$ & $674\pm4$ & $1.58\pm0.04$ & $0.5\pm0.02$ & $1.12$ & $-0.17$ & 92\\
\midrule
Host star & $M_{\rm BD}sini$ & $P_{\rm orb}$ & $a$ & $e$ & $M_*$ & [Fe/H] & Ref. \\
          & [$\rm M_{Jupiter}$] & [yr]  & [AU] &  & [$\rm M_{\odot}$] & [dex] & \\
\midrule
HAT-P-2  & $10.7^{+5.2}_{-2.2}$ & $23.272^{+7.1184}_{-4.1068}$ & $9^{+1.77}_{-1.08}$ & $0.37^{+0.13}_{-0.12}$ & $1.34$ & $0.04$ & 110\\
HD 106270  & $11\pm0.8$ & $7.9124\pm1.0678$ & $4.3\pm0.4$ & $0.402\pm0.054$ & $1.32$ & $0.08$ & 111\\
Kepler-407 A & $11.089\pm0.0493$ & $5.7269\pm0.013503$ & $3.27$ & $0.221\pm0.0003$ & $0.99$ & $0.35$ & 112\\
HATS-59  & $12.7\pm0.9$ & $3.8932\pm0.03833$ & $2.5\pm0.035$ & $0.08^{+0}_{-0.08}$ & $1.038$ & $0.18$ & 113\\
HD 219077  & $13.4\pm0.78$ & $15.094^{+0.13689}_{-0.1232}$ & $7.03\pm0.2$ & $0.768\pm0.004$ & $1.05$ & $-0.13$ & 114\\
HD 8673 A & $14.2\pm1.6$ & $4.4736\pm0.046543$ & $3.02\pm0.15$ & $0.723\pm0.016$ & $1.35$ & $0.14$ & 115\\
WASP-34  & $14.96^{+6.29}_{-3.39}$ & $11.206^{+2.0534}_{-1.4237}$ & $5.05^{+0.65}_{-0.46}$ & - & $1.01$ & $-0.02$ & 116\\
HD 219828  & $15.1\pm0.85$ & $13.117\pm0.20534$ & $5.96$ & $0.8118\pm0.0032$ & $1.24$ & $0.19$ & 117\\
CoRoT-20  & $17\pm1$ & $4.5859^{+0.052019}_{-0.046543}$ & $2.9\pm0.07$ & $0.6\pm0.03$ & $1.14$ & $0.14$ & 118\\
$^*$1SWASP J1407 & $20\pm6$ & $10.198\pm2.4641$ & $3.9\pm1.7$ & - & $0.9$ & - & 119\\
BEBOP-4 (AB) & $20.9\pm1.3$ & $4.9925^{+0.013963}_{-0.013415}$ & $3.634\pm0.011$ & $0.428\pm0.016$ & $1.966$ & $0.261$ & 120\\
HD 209262  & $32.3^{+1.6}_{-1.5}$ & $14.867^{+0.3833}_{-0.27379}$ & $6.16^{+0.15}_{-0.13}$ & $0.35\pm0.01$ & $1.03$ & $0.13$ & 121\\
HD 18757 A & $35.2\pm1.2$ & $108.93^{+17.988}_{-15.989}$ & $22\pm2.3$ & $0.943\pm0.007$ & $1.02$ & $-0.283$ & 121\\
BD+63 974  & $48\pm1.2$ & $5.8771\pm0.021355$ & $3.75\pm0.04$ & $0.209\pm0.006$ & $1.53$ & $-0.01$ & 122\\
V1128 Tau (AB) & $50$ & $16.099$ & - & $0.481$ & $1.67$ & - & 123\\
HD 8765  & $51.5^{+24.2}_{-8.5}$ & $7.6847^{+10.404}_{-1.389}$ & $4^{+3.1}_{-0.5}$ & $0.54$ & $1.03$ & $0.16$ & 124\\
HD 29587  & $55.2\pm9.2$ & $4.0548\pm0.060233$ & $2.597$ & $0.713\pm0.006$ & $1.01$ & $-0.527$ & 69\\
HD 72780  & $55.3^{+5.1}_{-2.9}$ & $17.488^{+24.183}_{-5.6961}$ & $7.3^{+5.7}_{-1.7}$ & $0.66$ & $1.22$ & $0.15$ & 124\\
$^*$WASP-81 & $56.6\pm0.2$ & $3.551^{+0.022177}_{-0.021355}$ & $2.426^{+0.044}_{-0.045}$ & $0.557\pm0.0044$ & $1.08$ & $-0.36$ & 125\\
\bottomrule
\multicolumn{8}{l}{$^*$ denote the transiting systems}\\
\multicolumn{8}{p{\textwidth}}{\textbf{Reference. }
87.\cite{Luhn19};
88.\cite{Lovis07};
89.\cite{Gonzalez17};
90.\cite{Li14};
91.\cite{Dai21};
92.\cite{Ma16};
93.\cite{Grieves17};
94.\cite{Winn10};
95.\cite{Lee23};
96.\cite{Almeida17};
97.\cite{Teng23};
98.\cite{Endl04};
99.\cite{Lee13};
100.\cite{Omiya09};
101.\cite{Kane09};
102.\cite{Ma13};
103.\cite{Barbato23};
104.\cite{Grandjean23};
105.\cite{Vaulato22};
106.\cite{Wilson16};
107.\cite{Diaz12};
108.\cite{Halbwachs12};
109.\cite{Latham02};
110.\cite{Beurs23};
111.\cite{Johnson11};
112.\cite{Weiss24};
113.\cite{Sarkis18};
114.\cite{Kane19};
115.\cite{Hartmann10};
116.\cite{Knutson14};
117.\cite{Santos16};
118.\cite{Rey18};
119.\cite{Kenworthy15};
120.\cite{Triaud25};
121.\cite{Bouchy16};
122.\cite{Niedzielski25};
123.\cite{Han24};
124.\cite{Patel07};
125.\cite{Triaud17}.
}
\end{longtable}

\end{appendix}

%






   
  



\end{document}